\documentclass{emulateapj}

\usepackage{amsmath,amssymb,verbatim}
\usepackage{lipsum}
\shorttitle{The First High-Resolution Spectroscopic Study of the Orphan Stream}
\shortauthors{Casey et al}

\begin{document}

\title{Hunting the Parent of the Orphan Stream II:\\The First High-Resolution Spectroscopic Study\altaffilmark{1}}

\author{Andrew R. Casey\altaffilmark{2}, Stefan C. Keller\altaffilmark{2}, Gary Da Costa\altaffilmark{2}, Anna Frebel\altaffilmark{3}, and Elizabeth Maunder\altaffilmark{2}}

\altaffiltext{1}{This paper includes data gathered with the 6.5 meter Magellan Telescopes located at Las Campanas Observatory, Chile.}
\altaffiltext{2}{Research School of Astronomy and Astrophysics, Australian National University, Canberra, ACT 2611, Australia; \email{andrew.casey@anu.edu.au}}
\altaffiltext{3}{Department of Physics, Massachusetts Institute of Technology \& Kavli Institute for Astrophysics and Space Research, Cambridge, MA 02139, USA}

\begin{abstract}
\noindent{}We present the first high-resolution spectroscopic study on the Orphan Stream for five stream candidates, observed with the Magellan Inamori Kyocera Echelle (MIKE) spectrograph on the Magellan Clay telescope. The targets were selected from the low-resolution catalog of \citet{casey;et-al_2013a}: 3 high-probability members, 1 medium and 1 low-probability stream candidate were observed. Our analysis indicates the low and medium-probability target are metal-rich field stars. The remaining three high-probability targets range over $\sim$1\,dex in metallicity, and are chemically distinct compared to the other 2 targets and all standard stars: low [$\alpha$/Fe] abundance ratios are observed, and lower limits are ascertained for [Ba/Y], which sit well above the Milky Way trend. These chemical signatures demonstrate that the undiscovered parent system is unequivocally a dwarf spheroidal galaxy, consistent with dynamical constraints inferred from the stream width and arc. As such, we firmly exclude the proposed association between NGC 2419 and the Orphan stream. A wide range in metallicities adds to the similarities between the Orphan stream and Segue 1, although low [$\alpha$/Fe] abundance ratios in the Orphan stream are in tension with high [$\alpha$/Fe] values observed in Segue 1. Open questions remain before Segue 1 could possibly be claimed as the `parent' of the Orphan stream. The parent system could well remain undiscovered in the southern sky.
\end{abstract}

\keywords{Galaxy: halo, structure --- Stars: Population II, late-type, abundances}

\section{Introduction}


A prolonged, quiescent period followed the creation of the Universe. Eventually, minuscule dark matter density perturbations initiated the collapse of dark matter, forming gravitational potential wells and giving rise to the condensation of primordial gas clouds \citep[e.g.,][]{els}. This process furnished the Universe with the earliest building blocks, laying the foundation for cosmological structure. Separate building blocks (i.e., gas fragments) underwent independent chemical enrichment before conglomerating to form larger systems. Evidence of this hierarchical formation \citep{searle;zinn_1978} remains observable today through the accretion of satellite systems onto the Milky Way halo. Although the formation history of the Galaxy is tangled and chaotic, it serves as an excellent -- and more importantly, accessible -- laboratory to investigate the evolution of the Universe since the earliest times.

Many stars in our Galaxy have formed in-situ, but a significant fraction have also been added through the accretion of smaller systems onto the Milky Way \citep[for a review see][]{helmi_2008}. As these satellite systems move in towards the Galaxy's gravitational potential, they are disrupted by tidal forces, causing loosely gravitationally bound stars to be strewn in forward and trailing directions as tidal tails or `stellar streams'. The observational evidence for ongoing accretion has significantly increased in the last decade since the discovery of the Sagittarius dwarf spheroidal undergoing tidal disruption \citep{ibata;et-al_1994,belokurov;et-al_2007,bell;et-al_2008,starkenburg;et-al_2008,deason;et-al_2012,drake;et-al_2013}.





Amongst the known substructures in the halo, the Orphan stream is particularly interesting. \citet{grillmair_2006} and \citet{Belokurov;et-al_2006} independently discovered the stream, spanning over $60^\circ$ in the sky from Ursa Major in the north to near Sextans in the south, where the SDSS coverage ends. The stream has properties that are unique from other halo substructures. It has an extremely low surface brightness, ranging from 32-40\,mag arcsec$^{-2}$, and a full width half-maximum of $\sim$2$^\circ$ on sky. At a distance of $\sim$20\,kpc \citep{belokurov;et-al_2007,grillmair_2006}, this corresponds to $\sim$700\,pc. The cross-section and luminosity of the stream are directly related to the mass and velocity dispersion of the parent satellite \citep{johnston_1998}. The Orphan stream width is significantly broader than every known globular cluster tidal tail \citep{odenkirchen;et-al_2003,grillmair_johnson_2006,grillmair_dionatos_2006a,grillmair_dionatos_2006b} and larger than the tidal diameter of all known globular clusters \citep[][2010 edition]{harris_1996}. As \citet{grillmair_2006} notes, if the cross-section of the stream is circular, then in a simple logarithmic potential with $v_c = 220$\,km s$^{-1}$ -- a reasonable first-order approximation for the Milky Way --  the expected random velocities of stars would be required to be $\sqrt{\langle\sigma_{v}^{2}\rangle} > 20$\,km s$^{-1}$ in order to produce the stream width. Such a velocity dispersion is significantly larger than the expected random motions of stars that have been weakly stripped from a globular cluster, implying that the Orphan stream's progenitor mass must be much larger than a classical globular cluster. Photometry indicates the stream is metal-poor, implying that negligible star formation has occurred since infall began several billion years ago. For a stream of this length to remain structurally coherent over such a long timescale, the progenitor is also likely to be dark-matter dominated. In their discovery papers, \citet{grillmair_2006} and \citet{belokurov;et-al_2007} concluded that the likely parent of the Orphan stream is a low-luminosity dwarf spheroidal (dSph) galaxy.

 
\citet{newberg;et-al_2010} were able to map the distance and velocity of the stream across the length of the SDSS catalogue using blue horizontal branch (BHB) and F-turnoff stars. They found the stream distance to vary between 19-47\,kpc, extending the 20-32\,kpc distance measurements made by \citet{belokurov;et-al_2007}, and has been extended to 55\,kpc by \citet{sesar;et-al_2013}. \citet{newberg;et-al_2010} note an increase in the density of the Orphan stream near the celestial equator $(l, b) = (253^\circ, 49^\circ)$, proposing the progenitor may be close to this position. The authors attempted to extend their trace of the stream using southern survey data \citep[e.g. SuperCOSMOS][]{supercosmos}, but to no avail; the stream's surface brightness is lower than the survey faint limit. It is still unclear whether the stream extends deep into the southern sky. \citet{newberg;et-al_2010} observe an increase in surface brightness near the celestial equator, but given the stream is closest near the celestial equator \citep{belokurov;et-al_2007}, an increase in stream density may be expected given a constant absolute magnitude. Nevertheless, with SDSS photometry and radial velocities from the SEGUE catalog, \citet{newberg;et-al_2010} were able to derive a prograde orbit with an eccentricity, apogalacticon and perigalacticon of $e = 0.7$, 90\,kpc and 16.4\,kpc respectively. The ability to accurately trace the Orphan stream to such extreme distances would make a powerful probe for measuring the galactic potential \citep[e.g. see][]{adrn_2013}. From their simulations, \citet{newberg;et-al_2010} find a halo and disk mass of $M(R < 60\,{\rm kpc}) = 2.6 \times 10^{11} M_\odot$, and similarly \citet{sesar;et-al_2013} find $2.7\times 10^{11} M_\odot$, which is $\sim$60\% lower than that found by \citet{xue;et-al_2008} and \citet{koposov;et-al_2008}, and slightly lower than the virial mass of $7 \times 10^{11} M_\odot$ derived by \citet{sales;et-al_2008}.
 
Metallicities derived from SEGUE medium-resolution spectra confirm photometric estimates indicating that the stream is metal-poor. A mean metallicity of [Fe/H]$ = -2.1 \pm 0.1$\,dex is found from BHB stars, with a range extending from $-$1.3 to $\sim-$3\,dex \citep{newberg;et-al_2010}. \citet{sesar;et-al_2013} also find a wide range in metallicities by tracing RR Lyrae stars along the stream: [Fe/H] = $-1.5$ to $-2.7$\,dex. However if F-turnoff stars from the \citet{newberg;et-al_2010} sample are included, the metallicity distribution function extends more metal-rich from $\sim-$3 to $-$0.5\,dex. 

The situation is complicated by interlopers and small number statistics, so the full shape of the metallicity distribution function (MDF) is unknown. To this end, \citet{casey;et-al_2013a} observed low-resolution spectroscopy for hundreds of stars towards the stream at the celestial equator. The authors targeted the less numerous K-giants \citep[where a mere 1.3 red giant branch stars are expected per square degree]{sales;et-al_2008,morrison_1993} and found a very weak detection of the stream from kinematics alone. Using wide selections in velocity, distance, proper motions, metallicities, and surface gravity, they identified nine highly likely Orphan stream giants. The velocity dispersion of their candidates is within the observational errors ($\sigma_{v} < 4$\,km s$^{-1}$), suggesting the stream is kinematically cold along the line-of-sight. Like \citet{newberg;et-al_2010} (and independently confirmed by \citet{sesar;et-al_2013}), they also found an extended range in metallicities: two stars below [Fe/H] $\leq-2.70$ and two stars near [Fe/H] $\sim -1.17$\,dex, all of which are consistent with stream membership. The mean metallicity of their sample was [Fe/H]$ = -1.63 \pm 0.19$\,dex, with a wide dispersion of $\sigma({\rm [Fe/H]}) = 0.56$\,dex. It appears the Orphan stream may have an extremely wide range in metallicity, consistent with the internal chemical enrichment typically observed in dwarf spheroidal (dSph) galaxies \citep{mateo_1998,tolstoy;et-al_2009,kirby;et-al_2011,frebel_2010}.


As the name suggests, the Orphan stream's parent satellite has yet to be found. In an effort to identify a progenitor, a number of systems have been identified as being plausibly associated with the Orphan stream. These include the linear Complex A H\,\textsc{i} clouds, as well as the globular clusters Palomar 1, Arp 2, Terzan 7 and Ruprecht 106. These systems were all identified along the great circle path of the stream. The low-luminosity dwarf satellites Segue 1 and Ursa Major II also lie along the great circle, although Segue 1 was considered an extended globular cluster until recently \citep{norris;et-al_2010,simon;et-al_2011}. 

\citet{belokurov;et-al_2007} first noted a possible association between the Orphan stream, Ursa Major II and Complex A. \citet{Jin;Lynden-Bell_2007} and \citet{fellhaur;et-al_2007} explored the possible association between the Orphan stream with Complex A and Ursa Major II, respectively. In the best-fitting Complex A model, the predicted heliocentric velocities did not match those found by \citet{belokurov;et-al_2007}, or later observations by \citet{newberg;et-al_2010}. The expected distances were also under-estimated by a factor of $\sim{}$3 when compared with observations, making the association with Complex A tenuous at best. In the Ursa Major II scenario, the stream's on-sky position was required to exactly overlap with a previous wrap, a somewhat contrived and unlikely scenario. \citet{newberg;et-al_2010} found that the Ursa Major II-Orphan stream connection was also not compelling, as the observed stream kinematics were not consistent with the Ursa Major II model.

Simulations involving the Complex A and Ursa Major II associations introduced an a priori assumption that the object (e.g. Ursa Major II or Complex A) was related to the Orphan stream, and consequently found an orbit to match. In contrast, \citet{sales;et-al_2008} approached the problem by fitting an orbit to a single wrap of the data, without assuming a parent satellite a priori. Their $N$-body simulations were inconsistent with either a Complex A or Ursa Major II association. Instead, the authors favor a progenitor with a luminosity $L \sim 2.3 \times 10^4 L_\odot$ or an absolute magnitude $M_r \sim -6.4$, consistent with the observation by \citet{belokurov;et-al_2007} of $M_r \sim -6.7$. Simulations by \citet{sales;et-al_2008} suggest the progenitor may be similar to the present-day `classical' Milky Way dwarfs like Carina, Draco, Leo II or Sculptor, but would be very close to being fully disrupted, which they suggest has occurred over the last 5.3 gigayear. Time of infall is a critical inference. Longer timescales produce streams that are too wide and diffuse, whereas shorter timescales do not reproduce the $\sim{}$60$^\circ$ stream length. The degeneracies between these simulation parameters are important to note, but in any case, there are robust conclusions that can be drawn irrespective of those degeneracies. For example, \citet{sales;et-al_2008} note that a globular cluster has a central density too high ($\sim10^{12} M_\odot$\,kpc$^{-3}$) to be fully disrupted along their Orphan stream orbit within a Hubble time. Given this constraint, and the lower limit on luminosity ($L > 2\times10^5 L_\odot$), a globular cluster progenitor seems unlikely from their models.

In addition to the work by \citet{sales;et-al_2008}, $N$-body simulations by \citet{newberg;et-al_2010} exclude all known halo globular clusters as possible progenitors. They conclude with the postulation of two possible scenarios: the progenitor is either an undiscovered satellite located between $(l, b) = (250^\circ, 50^\circ)$ and $(270^\circ, 40^\circ)$, or Segue 1 is the parent system. Segue 1, an ultrafaint dwarf galaxy \citep{simon;et-al_2011,norris;et-al_2010}, resides at a similar distance ($\sim$23\,kpc) along the great circle path of the Orphan stream. Segue 1 also shares velocities that are consistent with the Orphan stream at its nearest point. Moreover, both Segue 1 and the Orphan stream exhibit low velocity dispersions: on the order of $\sim$4\,km s$^{-1}$. The similarities in position, distance, and velocities between the two systems are indeed striking. However, it is not the only system alleged to be associated with the Orphan stream.


Given the extended apogalaction of 90\,kpc in the stream orbit found by \citet{newberg;et-al_2010}, \citet{bruns;kroupa_2011} reasoned the stream may be the tidal tail of the massive globular cluster NGC 2419. This system is the most distant and luminous outer halo globular cluster known ($\sim$85\,kpc, $M_V = -9.6$). It is unlike any other globular cluster in the Milky Way. Spectroscopic studies have confirmed photometric observations by \citet{photometry_for_ngc2419} that the system is metal-poor ([Fe/H] $= -2.15$), and identified a remarkable anti-correlation between Mg and K abundances \citep{cohen;et-al_2011,mucciarelli;et-al_2012}. The level of magnesium depletion ([Mg/Fe] $> -1.40$) is not seen anywhere else in the Galaxy or its satellite systems, neither is the enhanced potassium enrichment (up to [K/Fe] = 2) at the opposite end of this peculiar Mg-K anti-correlation. If NGC 2419 is the parent of the Orphan stream, then this unprecedented chemical signature ought to exist in disrupted stream members as an example of chemical tagging \citep[e.g., see][]{freeman;bland-hawthorn_2002,de_silva;et-al_2007,wylie-de-boer;et-al_2010,majewski;et-al_2012}.

\begin{deluxetable*}{lccccccccccc}
\tablecolumns{2}
\tabletypesize{\scriptsize}
\tablecaption{Observations\label{tab:observations}}
\tablehead{
	\colhead{Object\tablenotemark{a}} &
	\colhead{$\alpha$} &
	\colhead{$\delta$} &
	\colhead{$V$\tablenotemark{b}} &
	\colhead{UT Date} &
	\colhead{UT Time} &
	\colhead{Airmass} &
	\colhead{Exp. Time} &
	\colhead{S/N\tablenotemark{c}} &
	\colhead{$V_{\mbox{hel}}$} &
	\colhead{$V_{\mbox{err}}$} \\
 & (J2000) & (J2000) & (mag) & & & & (secs) & (px$^{-1}$) & (km s$^{-1}$) & (km s$^{-1}$)
}
\startdata
HD 41667		& 06:05:03.7	& $-$32:59:36.8	& 8.52	& 2011-03-13	& 23:40:52	& 1.01	& 90			& 272	& 297.8 	& 1.0 \\
HD 44007		& 06:18:48.6	& $-$14:50:44.2	& 8.06	& 2011-03-13	& 23:52:18	& 1.03	& 30 		& 239	& 163.4 	& 1.3 \\
HD 76932		& 08:58:44.2	& $-$16:07:54.2	& 5.86	& 2011-03-14	& 00:16:47	& 1.16	& 30 		& 289	& 119.2 	& 1.2 \\
HD 122563		& 14:02:31.8	& $+$09:41:09.9	& 6.02 	& 2011-03-13	& 07:15:04	& 1.28	& 30			& 230	& --23.4	& 1.0 \\
HD 136316		& 15:22:17.2	& $-$53:14:13.9	& 7.65	& 2011-03-14	& 09:37:26	& 1.12	& 90	 		& 335	& --38.2	& 1.1 \\
HD 141531		& 15:49:16.9	& $+$09:36:42.5	& 9.08	& 2011-03-14	& 09:52:00	& 1.31	& 90	 		& 280	& 2.6 		& 1.0 \\
HD 142948		& 16:00:01.6	& $-$53:51:04.1	& 8.03	& 2011-03-14	& 09:45:12	& 1.11	& 90	 		& 271	& 30.3 	& 0.9 \\
OSS 3 (L)		& 10:46:50.6	& $-$00:13:17.9	& 17.33 	& 2011-03-14	& 01:51:07	& 1.36	&4$\times$2500	& 48	& 217.9	& 1.0 \\ 
OSS 6 (H)		& 10:47:17.8	& $+$00:25:06.9	& 16.09 	& 2011-03-14	& 00:25:37	& 2.00	&3$\times$1600	& 59	& 221.2	& 1.0 \\ 
OSS 8 (H)		& 10:47:30.3	& $-$00:01:22.6	& 17.25	& 2011-03-14	& 04:44:04	& 1.16	&5$\times$1900	& 49	& 225.9	& 1.0 \\ 
OSS 14 (H)	& 10:49:08.3	& $+$00:01:59.3	& 16.27 	& 2011-03-15	& 00:32:17	& 1.88	&4$\times$1400	& 48 	& 225.1	& 1.0 \\ 
OSS 18 (M)	& 10:50:33.7	& $+$00:12:18.3	& 17.82 	& 2011-03-15	& 02:12:46	& 1.30	&4$\times$2100	& 31 	& 247.8	& 1.2
\enddata
\tablenotetext{a}{Probability of membership (Low, Medium, High) listed for Orphan stream candidates as defined by \citet{casey;et-al_2013a}.}
\tablenotetext{b}{$V$--band magnitudes for Orphan stream targets are estimated to be equivalent as $g$--band magnitudes.}
\tablenotetext{c}{S/N measured per pixel ($\sim$0.09\,{\AA} px$^{-1}$) at 600\,nm for each target.}
\end{deluxetable*}

In this study, we present an analysis of high-resolution spectroscopic observations for five Orphan stream candidates. The observations and data reduction are outlined in Section \ref{sec:observations}. In Section \ref{sec:analysis} we describe the details of our analysis to infer stellar parameters and chemical abundances. We discuss our results in Section \ref{sec:discussion}, including the implications for association between the Orphan stream and the two currently alleged stream progenitors: Segue 1 and NGC 2419. Finally, we conclude in Section \ref{sec:conclusions} with a summary of our findings.

\section{Observations and Data Reduction}
\label{sec:observations}


High-resolution spectra for five Orphan stream candidates and seven well-studied standard stars have been obtained with the Magellan Inamori Kyocera Echelle (MIKE) spectrograph \citep{bernstein;et-al_2003} on the Magellan Clay telescope. These objects were observed in March 2011 using a 1\arcsec{} wide slit in mean seeing of 0.9\arcsec. This slit configuration provides a continuous spectral coverage from 333\,nm to 915\,nm, with a spectral resolution of $\mathcal{R} = 25,000$ in the blue arm and $\mathcal{R} = 28,000$ in the red arm. A minimum of 10 exposures of each calibration type (biases, flat fields, and diffuse flats) were observed in the afternoon of each day, with additional flat-field and Th-Ar arc lamp exposures performed throughout the night to ensure an accurate wavelength calibration. The details of our observations are tabulated in Table \ref{tab:observations}. The signal-to-noise ($S/N$) ratio for the standard stars exceeds 200 per pixel, and varies between 30-60 for the Orphan stream candidates.

The candidates were chosen from the low-resolution spectroscopic study of \citet{casey;et-al_2013a}. From their classification, three of the selected stream candidates have high probability of membership to the Orphan stream. One target was classified with medium probability, and another with a low probability of membership. 

Initially we planned to observe many more high-priority targets. However, after our last exposure of target star OSS 18, inclement weather forced us to relinquish the remainder of our observing time. The data were reduced using the CarPy pipeline written by D. Kelson\footnote{http://obs.carnegiescience.edu/Code/mike}. Every reduced echelle aperture was carefully normalised using cubic splines with defined knot spacing. Extracted apertures were stacked together and weighted by their inverse variance to provide a continuous normalised spectrum for each object.

\section{Analysis}
\label{sec:analysis}
Each normalised, stitched spectrum was cross-correlated against a synthetic template to measure the radial velocity of each star. This was performed using a Python\footnote{http://www.python.org} implementation of the \citet{Tonry;Davis_1978} method. The wavelength region employed was from $845 \leq \lambda \leq 870$\,nm, and a synthetic spectrum of a metal-poor giant was used as the rest template. Heliocentric corrections have been applied to our radial velocity measurements, and the resultant heliocentric velocities are shown in Table \ref{tab:observations}.


Atomic data for absorption lines has been taken from \citet{yong;et-al_2005}, and these data are listed with their measured equivalent widths (EWs) in Table \ref{tab:atomic-data}. EWs for all atomic transitions were measured using the automatic profile fitting algorithm described in \citet{casey;et-al_2013b}. While this technique is accurate, and robust against blended lines as well as strong changes in the local continuum \citep{frebel;et-al_2013a,casey;et-al_2013b}, every fitted profile was visually inspected for quality. Spurious or false-positive measurements were removed and of order $\sim$5 EW measurements (of 528 atomic transitions in our line list\footnote{Although our final line list includes atomic data for 528 transitions, we have omitted data for 17 transitions from Table \ref{tab:atomic-data} as they were not significantly detected in any stars. See Section \ref{sec:chemical-abundances} for details.}) were manually re-measured for each object. We excluded transitions with reduced equivalent widths (REW), $\log(\textrm{EW}/\lambda) > -4.5$, in order to avoid using lines near the flat region of the curve-of-growth.

Given the range in metallicities for the Orphan stream candidates (see Section \ref{sec:analysis-metallicity}), these restrictions resulted in only 14 Fe\,\textsc{i} and 4 Fe\,\textsc{ii} acceptable transitions for our most metal-poor candidate, OSS 4. With so few lines available, minute changes in stellar parameters caused large trends and variations in the Fe line abundances. This resulted in a poor solution while performing the excitation and ionization equilibria. At this point, we opted to supplement our line list with transitions from \citet{roederer;et-al_2010}. Each additional transition was inspected in our most metal-rich candidate (OSS 18) to ensure that it was not blended with other features. Blended transitions were not added. As a result, the minimum number of acceptable Fe\,\textsc{i} and Fe\,\textsc{ii} transitions for any star increased to 48 and 12, respectively.

\setlength{\tabcolsep}{3.5pt}
\begin{deluxetable*}{clcrcccccccccccccc}[t!]
\tabletypesize{\scriptsize}
\tablecolumns{2}\tablecaption{Equivalent Widths for Standard and Program Stars\label{tab:atomic-data}}
\tablehead{
    \colhead{Wavelength} &
    \colhead{Species} &
    \colhead{$\chi$} &
    \colhead{$\log{gf}$} &
    \colhead{HD 41667} &
    \colhead{HD 44007} &
    \colhead{HD 76932} &
    \colhead{HD 122563} &
    \colhead{HD 136316} &
    \colhead{HD 141531} &
    \colhead{HD 142948} &
    \colhead{OSS 3} \\
 (nm) & & (eV) & & (m{\AA}) & (m{\AA}) & (m{\AA}) & (m{\AA}) & (m{\AA}) & (m{\AA}) & (m{\AA}) & (m{\AA}) &
}
\startdata
630.030	& O \textsc{i}		& 0.00	& --9.72	& \nodata	& 14.8	& \nodata	& 6.8	& \nodata	& 38.2	& 35.4	&\nodata	\\
636.378	& O \textsc{i}		& 0.02	&--10.16	& 11.6	& \nodata	& \nodata	& \nodata	& 11.6	& 14.7	& 12.7	&\nodata	\\
568.819	& Na \textsc{i}	& 2.11	& --0.42	& 71.9	& 30.9	& 57.4	& \nodata	& 29.3	& 38.1	& \nodata	& 55.2	\\
615.423	& Na \textsc{i}	& 2.10	& --1.53	& 10.7	& \nodata	& 7.8	& \nodata	& \nodata	& 5.2	& 28.4	&\nodata	\\
616.075	& Na \textsc{i}	& 2.10	& --1.23	& 16.6	& \nodata	& 13.4	& \nodata	& 4.6	& 7.0	& 43.4	& 17.4
\enddata
\tablenotetext{}{Table \ref{tab:atomic-data} is published in its entirety in the electronic edition. A portion is shown here for guidance regarding its form and content.}
\end{deluxetable*}

\subsection{Stellar Parameters}
We  employed the 1D plane-parallel model atmospheres of \citet{castelli;kurucz_2004} to infer stellar parameters. These $\alpha$-enhanced models assume that absorption lines form under local thermal equilibrium (LTE), ignore convective overshoot and any center-to-limb variations. We have interpolated within a grid of these model atmospheres following the prescription in \citet{casey;et-al_2013b}.

\subsubsection{Effective Temperature, $T_{\rm eff}$}
Effective temperatures for all stars have been found by excitation balance of neutral iron lines.  Since all EW measurements were visually inspected, we generally identified no outlier measurements during this stage. The most number of Fe\,\textsc{i} outliers ($>3\sigma$) removed while determining the effective temperature for any star was three. For each iteration in temperature, a linear fit was made to the data ($\chi$, $\log\epsilon$(Fe\,\textsc{i})). This slope was minimized with successive iterations of effective temperature. Final slopes less than $|10^{-3}|$\,dex eV$^{-1}$ were considered to be converged. 

The effective temperatures of our standard stars are in excellent agreement with the literature. On average, our effective temperatures are 13\,K cooler than the references listed in Table \ref{tab:stellar-parameters}. The largest discrepancy exists for HD 122563, a cool metal-poor giant. For extremely metal-poor stars, effective temperatures found through excitation balance are known to produce systematically cooler temperatures than those deduced by other methods \citep[e.g.][]{frebel;et-al_2013a}. We have chosen to remain consistent with the excitation balance approach, and accept the systematically cooler temperature of 4358\,K. The other noteworthy temperature deviant is HD 142948, where we find a temperature 337\,K hotter than that found by \citet{gratton;et-al_2000}. The reason for this discrepancy is not obvious.

\subsubsection{Microturbulence, $\xi$}
Microturbulence is necessary in 1D model atmospheres to represent large scale, 3D turbulent motions. The correct microturbulence will ensure that lines of the same species that are formed in deep and shallow photospheric depths will yield the same abundance. We have solved for the microturbulence by demanding a zero-trend in REW and abundance for all neutral iron lines. The resultant gradient between REW and abundance is typically $<|0.001|$\,dex, and the largest slope in any star is --0.004\,dex. 


\subsubsection{Surface Gravity, $\log{g}$}
The surface gravity for all stars has been inferred through the ionization balance of neutral and single ionized Fe lines. We iterated on surface gravity until the mean Fe\,\textsc{i} abundance matched the mean Fe\,\textsc{ii} abundance to within 0.01\,dex. This process was performed in concert while solving for all other stellar parameters. As a consequence of a systematically cooler temperature in HD 122563, we have obtained a somewhat lower surface gravity for this star, such that it sits above a metal-poor isochrone on a Hertzsprung-Russell diagram. Modulo HD 122563, the surface gravities for our standard stars are in good agreement with the literature: a mean offset of --0.22\,dex is observed (our $\log{g}$ values are lower). This difference remains within the mutual 1$\sigma$ uncertainties.

\begin{deluxetable*}{lcccccccccccl}[h!]
\tablecolumns{2}
\tabletypesize{\scriptsize}
\tablecaption{Stellar Parameters\label{tab:stellar-parameters}}
\tablehead{
& \multicolumn{4}{c}{\textbf{This Study}} &\multicolumn{2}{c}{ } & \multicolumn{5}{c}{\textbf{Literature}} \\
\cline{2-5} \cline{8-13}
	\colhead{Object} &
	\colhead{$T_{\rm eff}$} &
	\colhead{$\log{g}$} &
	\colhead{$\xi$} &
	\colhead{[Fe/H]} &\multicolumn{2}{c}{ }&
	\colhead{$T_{\rm eff}$} &
	\colhead{$\log{g}$} &
	\colhead{$\xi$} &
	\colhead{[Fe/H]} &
	\colhead{Reference} \\
	& (K) & (dex) & (km\,s$^{-1}$) & (dex) &\multicolumn{2}{c}{ }& (K) & (dex) & (km\,s$^{-1}$) & (dex) 
}
\startdata
\\
\multicolumn{12}{c}{\textbf{Standard Stars}} \\
\hline
HD 41667		& 4643	& 1.54	& 1.81	&  --1.18 &&& 4605 & 1.88 & 1.44 		& --1.16 & \citet{gratton;et-al_2000} 		\\ 
HD 44007		& 4820	& 1.66	& 1.70	&  --1.69 &&& 4850 & 2.00 & 2.20 		& --1.71 & \citet{fulbright_2000} 		\\ 
HD 76932		& 5835	& 3.93	& 1.42	&  --0.95 &&& 5849 & 4.11 & \nodata	& --0.88 & \citet{nissen;et-al_2000} 		\\ 
HD 122563		& 4358	& 0.14	& 2.77	&  --2.90 &&& 4843 & 1.62 & 1.80 		& --2.54 & \citet{yong;et-al_2013} 		\\ 
HD 136316		& 4347	& 0.44	& 2.15	&  --1.93 &&& 4414 & 0.94 & 1.70 		& --1.90 & \citet{gratton_sneden_1991} 	\\ 
HD 141531		& 4373	& 0.52	& 2.05	&  --1.65 &&& 4280 & 0.70 & 1.60 		& --1.68 & \citet{shetrone_1996} 		\\ 
HD 142948		& 5050	& 2.39	& 1.83	&  --0.64 &&& 4713 & 2.17 & 1.38 		& --0.77 & \citet{gratton;et-al_2000} 		\\ 
\hline
\\
\multicolumn{12}{c}{\textbf{Orphan Stream Candidates}} \\
\hline
OSS 3 (L)		& 5225	& 3.16	& 1.10	&  --0.86 &&& \nodata	& \nodata	& \nodata	&  --1.31	& \citet{casey;et-al_2013a}\\
OSS 6 (H)		& 4554	& 0.70	& 2.00	&  --1.75 &&& \nodata	& \nodata	& \nodata	&  --1.84	& \citet{casey;et-al_2013a}\\
OSS 8 (H)		& 4880	& 1.71	& 1.86	&  --1.62 &&& \nodata	& \nodata	& \nodata	&  --1.62	& \citet{casey;et-al_2013a}\\
OSS 14 (H)		& 4675	& 1.00	& 2.53	&  --2.66 &&& \nodata	& \nodata	& \nodata	&  --2.70	& \citet{casey;et-al_2013a}\\
OSS 18 (M)		& 5205	& 2.91	& 1.88	&  --0.62 &&& \nodata	& \nodata	& \nodata	&  --0.90	& \citet{casey;et-al_2013a}
\enddata
\end{deluxetable*}





\subsubsection{Metallicity, ${\rm [M/H]}$}
\label{sec:analysis-metallicity}

The final stage of iterative stellar parameter analysis is to derive the total metallicity. For these analyses we adopt the mean [Fe\,\textsc{i}/H] abundance as the overall metallicity [M/H]. A difference of ${|\textrm{[M/H]} - \langle\textrm{[Fe\,\textsc{i}/H]}\rangle| \leq 0.01}$\,dex was considered an acceptable model for the data. Excluding HD 122563, the mean metallicity difference for all other standard stars is 0.03\,dex less than values in the literature, with a standard deviation of 0.05\,dex. When the entire standard sample is included, these values change to --0.07$\pm$0.13\,dex. Acknowledging that our systematically low temperature for HD 122563 has resulted in differing stellar parameters, the rest of our standard stars exhibit excellent agreement with previous studies. We note that this effect is likely significantly smaller for the metal-poor star OSS 14, as this star is further down the giant branch with an effective temperature $\sim$300\,K hotter than HD 122563. Thus, we can be confident in the metallicity determination for OSS 14.

The offsets in metallicities between the values we derive from high-resolution spectroscopy and those found by \citet{casey;et-al_2013a} from low-resolution spectroscopy are noticeable. For the high probability targets (OSS 6, 8 and 14) the agreement is excellent: $\Delta$[Fe/H] = +0.04, 0.0, +0.09\,dex, respectively. We note that the uncertainties adopted in \citet{casey;et-al_2013a} are of the order $\pm0.3$\,dex; the discrepancies are well within the 1$\sigma$ uncertainties of {\it either} study. The largest difference between this study and that of \citet{casey;et-al_2013a} is in the lowest probability target (OSS 3), where we find a metallicity that is $+$0.45\,dex higher. 

It is reassuring that the metallicities of our high probability stars show the best agreement with the low-resolution measurements. Candidates were classified by \citet{casey;et-al_2013a} to have a low, medium, or high probability of membership with the Orphan stream. This classification was dependent on a number of observables, including metallicity. However, the metallicity determinations by \citet{casey;et-al_2013a} were calculated with the implied assumption that these stars were at a distance of $\sim$20\,kpc. The fact that our metallicities from high-resolution spectra are in excellent agreement with these initial metallicities indicates that our initial assumption was correct, and these high probability targets are at the approximate distance to the Orphan stream.\footnote{We note that to avoid bias from the low-resolution work of \citet{casey;et-al_2013a}, these high-resolution analyses took place with the information headers removed from each spectrum, and filenames were re-named to random strings. Original filenames were cross-matched only after the analysis was complete.}

\subsection{Uncertainties in Stellar Parameters}
\label{sec:stellar-parameter-uncertainties}
During excitation and ionization equilibria assessments, each fitted slope has an associated uncertainty due to the scatter in iron abundances. We have varied the effective temperature and microturbulence to match the formal uncertainty in each Fe\,\textsc{i} slope. Although these parameters are correlated, we have independently varied each parameter to reproduce the slope uncertainty. 

In order to estimate the uncertainty in $\log{g}$, the surface gravity has been adjusted until the mean difference in Fe\,\textsc{i} and Fe\,\textsc{ii} matches the quadrature sum of the variance in Fe\,\textsc{i} and Fe\,\textsc{ii} abundances. These uncertainties are listed in Table \ref{tab:stellar-parameter-uncertainties} for all standard and program stars. As these uncertainties are calculated with the assumption that they are uncorrelated, they are therefore possibly underestimated. As such we have assumed a minimum uncertainty of 125\,K in $T_{\rm eff}$, 0.3\,dex in $\log{g}$ and 0.2\,km s$^{-1}$ in $\xi$. Thus, the adopted uncertainty for each star is taken as the maximum of these values and those listed in Table \ref{tab:stellar-parameter-uncertainties}.\\

\begin{deluxetable}{lcccccccccccccl}[h!]
\tablecolumns{1}
\tabletypesize{\scriptsize}
\tablecaption{Uncorrelated Uncertainties in Stellar Parameters\label{tab:stellar-parameter-uncertainties}}
\tablehead{
	\colhead{Object} &
	\colhead{$\sigma(T_{\rm eff})$} &
	\colhead{$\sigma(\log{g})$} &
	\colhead{$\sigma(\xi)$} \\
	& (K) & (dex) & (km\,s$^{-1}$)
}
\startdata
HD 41667	&	47 &	0.08 &	0.05 \\
HD 44007	&	51 &	0.08 &	0.17 \\
HD 76932	&	61 &	0.10 &	0.14 \\
HD 122563	&	15 &	0.01 &	0.07 \\
HD 136316	&	34 &	0.02 &	0.05 \\
HD 141531	&	40 &	0.03 &	0.05 \\
HD 142948	&	62 &	0.12 &	0.09 \\
OSS 3 (L)	&	73 &	0.08 &	0.11 \\
OSS 6 (H)	&	39 & 0.04 &	0.15 \\
OSS 8 (H)	&	74 & 0.17 &	0.23 \\
OSS 14 (H)&	51 & 0.05 &	0.13 \\
OSS 18 (M)&	135& 0.15 &	0.28
\enddata
\end{deluxetable}

\subsection{Chemical Abundances}
\label{sec:chemical-abundances}
Chemical abundances are described following the standard nomenclature\footnote{$\log\epsilon(X) = \log{\left(\frac{N_X}{N_H}\right)} + 12$}, and comparisons made with reference to the Sun\footnote{$\left[\frac{X}{H}\right] = \log\epsilon{\left(\frac{N_X}{N_H}\right)}_{\rm Star} - \log\epsilon{\left(\frac{N_X}{N_H}\right)}_{\rm Sun}$} have been calculated using the solar composition described in \citet{asplund;et-al_2009}. We note that some transitions have been removed from our line list (Table \ref{tab:atomic-data}) when they were not detected in any star (e.g., Sc\,\textsc{i}, Zr\,\textsc{i}, Zr\,\textsc{ii}, Rb\,\textsc{i}, and Ce\,\textsc{ii} transitions).

\subsubsection{Carbon}

Carbon has been measured from synthesizing the 431.3\,nm and 432.3\,nm CH features. The best fitting abundances inferred from these two syntheses were within 0.05\,dex for every star. Nevertheless, have adopted a conservative total uncertainty of $\pm$0.15\,dex for these measurements. The mid-point of these two measurements is listed in Table \ref{tab:chemical-abundances}. Only the 432.3\,nm region was synthesized for OSS 18, as significant absorption of atomic lines was present in the bluer band head, and we deemed the 432.3\,nm region to yield a more precise determination of carbon abundance.

\subsubsection{Light Odd-Z Elements (Na, Al, K)}
Sodium, aluminium and potassium are primarily produced through carbon, neon and oxygen burning in massive stars, before they are ejected into the interstellar medium \citep{woosley;weaver_1995}. 

Although our line list includes three clean, unblended Na lines, not all were detectable in the Orphan stream candidates. Generally only one Na and Al line was available for the Orphan stream stars. In these cases, a minimum standard deviation of 0.10\,dex has been assumed when calculating abundance uncertainties (see Section \ref{sec:chemical-abundance-uncertainties}). Wherever there were no detectable lines for a given element, an upper limit was determined by synthesis of the strongest transition in our line list. At least one K line was detected in every standard and program star. These lines at 766\,nm and 769\,nm are quite strong, but fall directly between a strong telluric band head. Usually both lines were detected, but one was dominated by the Earth's atmospheric absorption. In these cases we rejected the contaminated line and adopted the single, unaffected K line.

\subsubsection{$\alpha$-elements (O, Mg, Ca, Si, Ti)}

The $\alpha$-elements (O, Mg, Ca, Si and Ti) are produced during hydrostatic burning of carbon, neon and silicon by $\alpha$-particle capture. Following Type II supernovae, the $\alpha$-enriched material is dispersed into the interstellar medium and contributes to the next generation of star formation. Although Ti is not formally an $\alpha$-element ($Z = 22$), it generally tracks with $\alpha$-elements, and has been included here to facilitate a comparison with the Local Group study of \citet{venn;et-al_2004}.

Generally, the Orphan stream candidates have lower $\alpha$-element abundances compared to iron than their halo counterparts. The situation is a little ambiguous for [O/Fe] compared to other $\alpha$-elements. We have employed the forbidden lines at 630\,nm and 636\,nm to measure oxygen abundances in our stars. Given the weakness of these lines, oxygen was immeasurable in all of the Orphan stream candidates. In place of abundances, upper limits have been determined from spectral synthesis of the region. The synthesis line list includes the Ni\,\textsc{i} feature hidden within the 630\,nm absorption profile \citep{allende-prieto;et-al_2001}. In the low $S/N$ ratio case of OSS 18, the forbidden oxygen region was sufficiently dominated by telluric absorption such that we deemed even a robust upper limit to be indeterminable.

\begin{figure}[h]
	\includegraphics[width=\columnwidth]{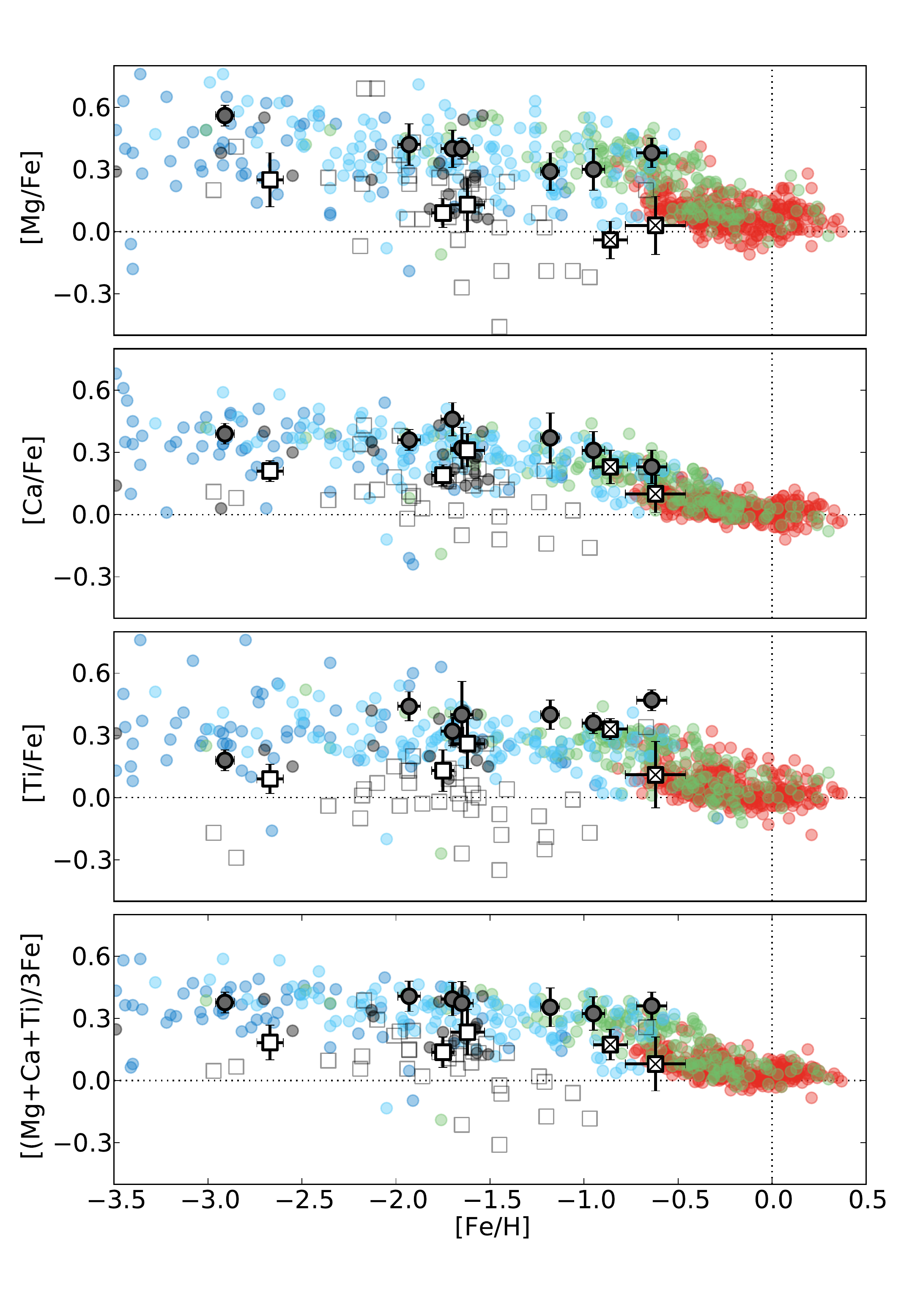}
	\caption{$\alpha$-element abundance ratios (Mg\,\textsc{i}, Ca\,\textsc{i}, Ti\,\textsc{ii} shown) for field standards (grey, filled), high-probability Orphan stream candidates (white squares), and lower-probability Orphan stream candidates (white squares with crosses) that are deemed to be interlopers. Data compiled by \citet{venn;et-al_2004,venn;et-al_2006} for the present-day dSph galaxies (grey squares) and the Milky Way is shown, using the same colour scheme adopted in \citet{venn;et-al_2004}: thin disk  (red), thick disk (green), halo (cyan), high velocity halo (dark blue), and retrograde halo stars (black).
	}
	\label{fig:alpha-fe}
\end{figure}

The [Mg/Fe] abundance ratios for the Orphan stream targets are noticeably lower than any of the other $\alpha$-elements. In fact, OSS 18 and 3 exhibit solar or sub-solar [Mg/Fe] abundances, and the mean [$\alpha$/Fe] abundances in the lower panel of Figure \ref{fig:alpha-fe} are most affected by [Mg/Fe]. Nevertheless, [Ca/Fe] and [Si/Fe] abundances for Orphan stream targets are systematically lower than their Milky Way counterparts. In Figure \ref{fig:alpha-fe}, Ti\,\textsc{ii} has been adopted for [Ti/Fe] as few reliable Ti\,\textsc{i} transitions were available for analysis.

Lower [$\alpha$/Fe] abundance ratios are less apparent for our lower probability Orphan stream candidates, OSS 18 and OSS 3. Although their mean [$\alpha$/Fe] $\sim$ +0.20\,dex, with [Fe/H] $\sim$ --0.70\,dex they are significantly more metal-rich than expected for the Orphan stream. Therefore their [$\alpha$/Fe] abundance ratios are consistent with the [$\alpha$/Fe]-[Fe/H] trend of the Milky Way. As discussed further in Section \ref{sec:stream-membership}, these low probability targets are unlikely to be true Orphan stream members, and are marked appropriately in Figure \ref{fig:alpha-fe}.

\subsubsection{Fe-peak elements (Sc, V, Cr, Mn, Co, Ni, Cu, Zn)}
The Fe-peak elements ($Z = 23$ to 30) are primarily produced by explosive nucleosynthesis during oxygen, neon, and silicon burning. The ignition of these burning phases occurs both from Type II SN of massive stars, and once an accreting white dwarf exceeds the Chandrasekhar limit, causing spontaneous ignition of carbon and an eventual Type Ia supernova. The abundance of the Fe-peak elements with respect to iron in the Milky Way are typically either flat (i.e., [X/Fe] $\sim$ 0; Cr\,\textsc{ii}, Ni) or trend positively (Sc, Cr\,\textsc{i}, Mn, Cu) with overall [Fe/H] \citep[e.g.,][]{ishigaki;et-al_2013,yong;et-al_2013}.

Scandium absorption profiles have appreciable broadening due to hyperfine splitting. As such, we have determined Sc abundances for all stars from spectral synthesis with hyperfine splitting taken into account. However, for OSS\,3 we found the broadening due to hyperfine structure to be negligible, and the variance in our synthesis measurements was larger than our abundance measurements. The mean Sc\,\textsc{ii} abundance to be $\log\epsilon({\rm Sc}) = 2.22 \pm 0.23$ from synthesis of 5 lines, where the EWs of 11 Sc\,\textsc{ii} lines yields $\log\epsilon({\rm Sc}) = 2.24 \pm 0.09$. Therefore in the case of OSS\,3, we have adopted the Sc\,\textsc{ii} abundances from EWs.

Other Fe-peak elements with hyperfine structure (e.g. V, Mn, Co, Cu) have been synthesised with the relevant isotopic and/or hyperfine splitting included. The random measurement scatter in V, Cr, and Mn abundances is typically low ($<$0.05\,dex) for all the standard and program stars. However for Co, there was a noticeable increase in line-to-line scatter for the Orphan stream stars compared to the standard stars. There is a factor of $\sim$6 difference in $S/N$ ratio between the two samples which can explain this variance. In the Orphan stream stars, no synthesised Co profiles were sufficiently `worse' than each other to qualify exclusion. The case was quite different for Ni, where a plethora of clean lines (without appreciable hyperfine structure) are available, and the uncertainties in $\log\epsilon({\rm Ni})$ abundances due to the uncertainties in stellar parameters generally cancel with $\log\epsilon({\rm Fe})$, yielding excellent measurements of [Ni/Fe]. Cu abundances and upper limits are derived from the synthesis of a single neutral Cu line at 510.5\,nm. The inclusion of hyperfine structure was most important for Cu, as a Cu abundance measured directly from an EW produced a systematically higher $\log\epsilon({\rm Cu})$, on the order of +0.4\,dex. Zn abundances were calculated directly from the EWs of the 472.2\,nm and 481.0\,nm transitions where available. All of our Fe-peak abundances with respect to iron are consistent with the observed chemical trends in the Milky Way \citep{ishigaki;et-al_2013,yong;et-al_2013}.

\subsubsection{Neutron-capture elements (Sr, Y, Ba, La, Nd, Eu)}

The atomic absorption lines of these heavy elements have appreciable broadening due to hyperfine structure and isotopic splitting. Where applicable, the relevant hyperfine and/or isotopic splitting has been employed, and solar isotopic compositions have been adopted.

Sr and Y are elements of the first $n$-capture peak. While only one line (421.5\,nm) was synthesised for Sr\,\textsc{ii}, generally three unblended Y lines were available. When no Y lines were detected above $3\sigma$, an upper limit was ascertained from the 520\,nm line -- the strongest in our line list. Sr and Y abundances generally agree with each other in our stars, with the exception of high probability stream candidates OSS 6 and OSS 8. In these cases only upper limits could be determined for Y\,\textsc{ii}, and those limits are $\sim$0.6\,dex lower than our measured Sr\,\textsc{ii} abundances. The remaining high probability candidate, OSS 14, also has an upper limit for Y\,\textsc{ii} ([Y/Fe] $< -0.40$) but this limit is much closer to our Sr\,\textsc{ii} measurement ([Sr/Fe] = $-0.42$). The Y upper limits for our high probability Orphan stream stars results in lower limits for [Ba/Y], which for OSS 6 and OSS 8, are well in excess of the Milky Way trend (Figure \ref{fig:ba-y}).

Lanthanum and neodymium abundances for our stars are consistent with the chemical evolution of the Milky Way. Europium, a $r$-process dominated element, has been measured by synthesising the 664.5\,nm Eu\,\textsc{ii} transition. Although the spread in [Eu/Fe] abundances for our stars is wide, no significant $r$-process enhancement is observed. 

\subsection{Chemical Abundance Uncertainties Due to Stellar Parameters}
\label{sec:chemical-abundance-uncertainties}
Although the standard error about the mean ($\sigma_{\bar{x}}$) abundance in Table \ref{tab:chemical-abundances} can be quite low, the uncertainties in stellar parameters will significantly contribute to the total error budget for any given abundance. Furthermore, the total uncertainty in abundance ratios (e.g., [A/B]) depends on how the uncertainties in elements A and B are correlated with stellar parameters. \\

\begin{longtable*}{lcrcrrcclcrcrrc}
\caption{Chemical abundances\label{tab:chemical-abundances}}
\\
\cline{1-15}
Species & N & $\log\epsilon(X)$ & $\sigma$ & [X/H] & [X/Fe] & $\sigma_{\bar{x}}$ && 
Species & N & $\log\epsilon(X)$ & $\sigma$ & [X/H] & [X/Fe] & $\sigma_{\bar{x}}$ \tabularnewline
\cline{1-15} \tabularnewline
\endfirsthead
\caption{Chemical abundances\label{tab:chemical-abundances}}
\\
\cline{1-15}
Species & N & $\log\epsilon(X)$ & $\sigma$ & [X/H] & [X/Fe] & $\sigma_{\bar{x}}$ && 
Species & N & $\log\epsilon(X)$ & $\sigma$ & [X/H] & [X/Fe] & $\sigma_{\bar{x}}$ \tabularnewline
\cline{1-15}
\endhead
\hline
\multicolumn{15}{r}{Continued..}
\endfoot
\hline
\endlastfoot

\\
\multicolumn{7}{c}{\textbf{HD 41667}} & \colhead{} & \multicolumn{7}{c}{\textbf{HD 44007}} \\
\cline{1-7} \cline{9-15}


   C (CH)       &   2 &    6.95 &    0.20 &  --1.48 &  --0.24 &    0.15 &&   C (CH)       &   2 &    6.73 &    0.20 &  --1.70 &  --0.01 &    0.15 \\
   O \textsc{I} &   1 &    7.85 & \nodata &  --0.84 &    0.34 & \nodata &&   O \textsc{I} &   1 &    7.43 & \nodata &  --1.26 &    0.44 & \nodata \\
  Na \textsc{I} &   3 &    4.95 &    0.10 &  --1.29 &  --0.10 &    0.06 &&  Na \textsc{I} &   1 &    4.57 & \nodata &  --1.67 &    0.03 & \nodata \\
  Mg \textsc{I} &   7 &    6.71 &    0.13 &  --0.89 &    0.29 &    0.05 &&  Mg \textsc{I} &   5 &    6.29 &    0.06 &  --1.31 &    0.40 &    0.03 \\
  Al \textsc{I} &   4 &    5.27 &    0.11 &  --1.18 &  --0.00 &    0.05 &&  Al \textsc{I} &   0 & \nodata & \nodata & \nodata & \nodata & \nodata \\
  Si \textsc{I} &   8 &    6.58 &    0.14 &  --0.93 &    0.26 &    0.05 &&  Si \textsc{I} &   7 &    6.18 &    0.09 &  --1.33 &    0.37 &    0.03 \\
   K \textsc{I} &   2 &    4.66 &    0.00 &  --0.38 &    0.81 &    0.00 &&   K \textsc{I} &   2 &    4.43 &    0.05 &  --0.60 &    1.10 &    0.04 \\
  Ca \textsc{I} &  18 &    5.52 &    0.07 &  --0.82 &    0.37 &    0.02 &&  Ca \textsc{I} &  20 &    5.10 &    0.09 &  --1.24 &    0.46 &    0.02 \\
 Sc \textsc{II} &  14 &    2.10 &    0.10 &  --1.05 &    0.13 &    0.03 && Sc \textsc{II} &  12 &    1.45 &    0.08 &  --1.70 &    0.00 &    0.02 \\
  Ti \textsc{I} &  25 &    3.90 &    0.16 &  --1.05 &    0.13 &    0.03 &&  Ti \textsc{I} &  20 &    3.35 &    0.08 &  --1.60 &    0.11 &    0.02 \\
 Ti \textsc{II} &  28 &    4.16 &    0.18 &  --0.79 &    0.40 &    0.03 && Ti \textsc{II} &  36 &    3.57 &    0.13 &  --1.38 &    0.32 &    0.02 \\
   V \textsc{I} &   4 &    2.82 &    0.09 &  --1.11 &    0.07 &    0.05 &&   V \textsc{I} &   4 &    2.24 &    0.05 &  --1.69 &    0.01 &    0.02 \\
  Cr \textsc{I} &  12 &    4.28 &    0.07 &  --1.36 &  --0.18 &    0.02 &&  Cr \textsc{I} &  15 &    3.75 &    0.06 &  --1.89 &  --0.19 &    0.02 \\
 Cr \textsc{II} &   3 &    4.65 &    0.13 &  --0.99 &    0.20 &    0.08 && Cr \textsc{II} &   3 &    4.10 &    0.04 &  --1.54 &    0.16 &    0.03 \\
  Mn \textsc{I} &   5 &    4.12 &    0.17 &  --1.31 &  --0.13 &    0.08 &&  Mn \textsc{I} &   7 &    3.30 &    0.10 &  --2.13 &  --0.43 &    0.04 \\
  Fe \textsc{I} &  72 &    6.32 &    0.11 &  --1.18 &    0.00 &    0.01 &&  Fe \textsc{I} &  74 &    5.80 &    0.12 &  --1.70 &    0.00 &    0.01 \\
 Fe \textsc{II} &  17 &    6.32 &    0.08 &  --1.18 &    0.00 &    0.02 && Fe \textsc{II} &  17 &    5.80 &    0.12 &  --1.70 &    0.01 &    0.03 \\
  Co \textsc{I} &   4 &    3.85 &    0.05 &  --1.14 &    0.04 &    0.03 &&  Co \textsc{I} &   3 &    3.29 &    0.09 &  --1.70 &    0.00 &    0.05 \\
  Ni \textsc{I} &  20 &    4.96 &    0.13 &  --1.26 &  --0.08 &    0.03 &&  Ni \textsc{I} &  23 &    4.46 &    0.12 &  --1.76 &  --0.05 &    0.03 \\
  Cu \textsc{I} &   1 &    2.86 & \nodata &  --1.33 &  --0.15 & \nodata &&  Cu \textsc{I} &   1 &    1.90 & \nodata &  --2.29 &  --0.59 & \nodata \\
  Zn \textsc{I} &   2 &    3.39 &    0.04 &  --1.17 &    0.01 &    0.03 &&  Zn \textsc{I} &   2 &    2.88 &    0.08 &  --1.68 &    0.02 &    0.06 \\
 Sr \textsc{II} &   1 &    1.75 & \nodata &  --1.12 &    0.06 & \nodata && Sr \textsc{II} &   1 &    1.19 & \nodata &  --1.68 &    0.02 & \nodata \\
  Y \textsc{II} &   3 &    1.08 &    0.12 &  --1.13 &    0.05 &    0.07 &&  Y \textsc{II} &   3 &    0.35 &    0.06 &  --1.86 &  --0.15 &    0.04 \\
 Ba \textsc{II} &   2 &    0.97 &    0.03 &  --1.21 &  --0.02 &    0.02 && Ba \textsc{II} &   2 &    0.43 &    0.08 &  --1.75 &  --0.04 &    0.05 \\
 La \textsc{II} &   2 &    0.14 &    0.03 &  --0.97 &    0.22 &    0.02 && La \textsc{II} &   2 &  --0.42 &    0.00 &  --1.52 &    0.18 &    0.00 \\
 Nd \textsc{II} &   6 &    0.56 &    0.06 &  --0.86 &    0.32 &    0.02 && Nd \textsc{II} &   7 &  --0.26 &    0.09 &  --1.68 &    0.03 &    0.03 \\
 Eu \textsc{II} &   1 &  --0.16 & \nodata &  --0.68 &    0.50 & \nodata && Eu \textsc{II} &   1 &  --1.16 & \nodata &  --1.68 &    0.02 & \nodata \\

\cline{1-7} \cline{9-15} \\ \\
\multicolumn{7}{c}{\textbf{HD 76932}} && \multicolumn{7}{c}{\textbf{HD 122563}} \\
\cline{1-7} \cline{9-15}

   C (CH)       &   2 &    7.63 &    0.20 &  --0.79 &    0.15 &    0.15 &&   C (CH)       &   2 &    5.26 &    0.20 &  --3.17 &  --0.27 &    0.15 \\
   O \textsc{I} &   1 & $<$8.33 & \nodata &$<$--0.36& $<$0.58 & \nodata &&   O \textsc{I} &   1 &    6.15 & \nodata &  --2.54 &    0.37 & \nodata \\
  Na \textsc{I} &   3 &    5.44 &    0.03 &  --0.80 &    0.15 &    0.02 &&  Na \textsc{I} &   1 & $<$3.48 & \nodata &$<$--2.76& $<$0.14 & \nodata \\
  Mg \textsc{I} &   8 &    6.94 &    0.18 &  --0.66 &    0.30 &    0.06 &&  Mg \textsc{I} &   8 &    5.25 &    0.08 &  --2.35 &    0.56 &    0.03 \\
  Al \textsc{I} &   4 &    5.56 &    0.08 &  --0.89 &    0.07 &    0.04 &&  Al \textsc{I} &   1 & $<$4.83 & \nodata &$<$--1.62& $<$1.28 & \nodata \\
  Si \textsc{I} &   9 &    6.80 &    0.18 &  --0.71 &    0.24 &    0.06 &&  Si \textsc{I} &   1 &    5.21 & \nodata &  --2.30 &    0.61 & \nodata \\
   K \textsc{I} &   2 &    4.96 &    0.06 &  --0.07 &    0.89 &    0.05 &&   K \textsc{I} &   1 &    2.75 & \nodata &  --2.28 &    0.63 & \nodata \\
  Ca \textsc{I} &  23 &    5.69 &    0.09 &  --0.65 &    0.31 &    0.02 &&  Ca \textsc{I} &  19 &    3.81 &    0.05 &  --2.53 &    0.39 &    0.01 \\
 Sc \textsc{II} &  16 &    2.38 &    0.10 &  --0.77 &    0.18 &    0.02 && Sc \textsc{II} &  15 &    0.17 &    0.07 &  --2.98 &  --0.07 &    0.02 \\
  Ti \textsc{I} &  17 &    4.16 &    0.06 &  --0.79 &    0.16 &    0.01 &&  Ti \textsc{I} &  17 &    2.06 &    0.04 &  --2.89 &    0.02 &    0.01 \\
 Ti \textsc{II} &  38 &    4.35 &    0.12 &  --0.60 &    0.36 &    0.02 && Ti \textsc{II} &  42 &    2.22 &    0.10 &  --2.73 &    0.18 &    0.02 \\
   V \textsc{I} &   4 &    3.28 &    0.12 &  --0.65 &    0.31 &    0.06 &&   V \textsc{I} &   1 &    0.81 & \nodata &  --3.12 &  --0.21 & \nodata \\
  Cr \textsc{I} &  18 &    4.58 &    0.07 &  --1.06 &  --0.11 &    0.02 &&  Cr \textsc{I} &  11 &    2.39 &    0.08 &  --3.25 &  --0.34 &    0.02 \\
 Cr \textsc{II} &   3 &    4.86 &    0.04 &  --0.78 &    0.18 &    0.02 && Cr \textsc{II} &   3 &    2.83 &    0.03 &  --2.81 &    0.10 &    0.02 \\
  Mn \textsc{I} &   9 &    4.31 &    0.10 &  --1.12 &  --0.17 &    0.03 &&  Mn \textsc{I} &   7 &    2.12 &    0.06 &  --3.31 &  --0.39 &    0.02 \\
  Fe \textsc{I} &  96 &    6.55 &    0.11 &  --0.95 &    0.00 &    0.01 &&  Fe \textsc{I} & 165 &    4.59 &    0.11 &  --2.91 &    0.00 &    0.01 \\
 Fe \textsc{II} &  20 &    6.55 &    0.13 &  --0.95 &    0.00 &    0.03 && Fe \textsc{II} &  22 &    4.60 &    0.11 &  --2.90 &    0.01 &    0.02 \\
  Co \textsc{I} &   4 &    4.09 &    0.12 &  --0.90 &    0.05 &    0.06 &&  Co \textsc{I} &   5 &    2.23 &    0.14 &  --2.76 &    0.15 &    0.06 \\
  Ni \textsc{I} &  25 &    5.32 &    0.11 &  --0.90 &    0.05 &    0.02 &&  Ni \textsc{I} &  19 &    3.48 &    0.10 &  --2.74 &    0.17 &    0.02 \\
  Cu \textsc{I} &   1 &    2.95 & \nodata &  --1.24 &  --0.29 & \nodata &&  Cu \textsc{I} &   1 & $<$0.10 & \nodata &$<$--4.19&$<$--1.19& \nodata \\
  Zn \textsc{I} &   2 &    3.71 &    0.02 &  --0.85 &    0.10 &    0.01 &&  Zn \textsc{I} &   2 &    1.83 &    0.06 &  --2.72 &    0.19 &    0.05 \\
 Sr \textsc{II} &   1 &    2.00 & \nodata &  --0.87 &    0.08 & \nodata && Sr \textsc{II} &   1 &  --0.64 & \nodata &  --3.51 &  --0.60 & \nodata \\
  Y \textsc{II} &   2 &    1.29 &    0.04 &  --0.92 &    0.03 &    0.02 &&  Y \textsc{II} &   1 &  --0.82 & \nodata &  --3.03 &  --0.12 & \nodata \\
 Ba \textsc{II} &   2 &    1.34 &    0.04 &  --0.84 &    0.12 &    0.03 && Ba \textsc{II} &   2 &  --1.94 &    0.06 &  --4.12 &  --1.20 &    0.04 \\
 La \textsc{II} &   1 &    0.69 & \nodata &  --0.41 &    0.54 & \nodata && La \textsc{II} &   1 &$<$--1.51& \nodata &$<$--2.61& $<$0.29 & \nodata \\
 Nd \textsc{II} &   3 &    0.81 &    0.11 &  --0.61 &    0.35 &    0.07 && Nd \textsc{II} &   1 &$<$--1.83& \nodata &$<$--3.25&$<$--0.35& \nodata \\
 Eu \textsc{II} &   1 &$<$--0.27& \nodata &$<$--0.79& $<$0.15 & \nodata && Eu \textsc{II} &   1 &$<$--1.72& \nodata &$<$--2.24& $<$0.66 & \nodata \\

\cline{1-7} \cline{9-15} \\ \\
\multicolumn{7}{c}{\textbf{HD 136316}} && \multicolumn{7}{c}{\textbf{HD 141531}} \\
\cline{1-7} \cline{9-15}

   C (CH)       &   2 &    6.05 &    0.20 &  --2.38 &  --0.47 &    0.15 &&   C (CH)       &   2 &    6.31 &    0.20 &  --2.12 &  --0.42 &    0.15 \\
   O \textsc{I} &   1 &    7.11 & \nodata &  --1.58 &    0.35 & \nodata &&   O \textsc{I} &   2 &    7.36 &    0.01 &  --1.33 &    0.31 &    0.00 \\
  Na \textsc{I} &   2 &    4.21 &    0.08 &  --2.03 &  --0.10 &    0.06 &&  Na \textsc{I} &   3 &    4.40 &    0.07 &  --1.84 &  --0.19 &    0.04 \\
  Mg \textsc{I} &   8 &    6.09 &    0.29 &  --1.51 &    0.42 &    0.10 &&  Mg \textsc{I} &   9 &    6.36 &    0.27 &  --1.24 &    0.40 &    0.09 \\
  Al \textsc{I} &   1 & $<$4.58 & \nodata &$<$--1.87& $<$0.04 & \nodata &&  Al \textsc{I} &   1 &    4.82 & \nodata &  --1.63 &    0.01 & \nodata \\
  Si \textsc{I} &   8 &    5.93 &    0.17 &  --1.58 &    0.35 &    0.06 &&  Si \textsc{I} &   7 &    6.08 &    0.15 &  --1.43 &    0.21 &    0.06 \\
   K \textsc{I} &   2 &    3.83 &    0.03 &  --1.21 &    0.72 &    0.02 &&   K \textsc{I} &   2 &    4.13 &    0.06 &  --0.90 &    0.74 &    0.04 \\
  Ca \textsc{I} &  19 &    4.77 &    0.08 &  --1.57 &    0.36 &    0.02 &&  Ca \textsc{I} &  17 &    5.02 &    0.06 &  --1.32 &    0.32 &    0.01 \\
 Sc \textsc{II} &  16 &    1.24 &    0.08 &  --1.91 &    0.01 &    0.02 && Sc \textsc{II} &  12 &    1.56 &    0.05 &  --1.59 &    0.05 &    0.01 \\
  Ti \textsc{I} &  26 &    3.02 &    0.13 &  --1.93 &  --0.01 &    0.03 &&  Ti \textsc{I} &  26 &    3.34 &    0.13 &  --1.61 &    0.03 &    0.03 \\
 Ti \textsc{II} &  39 &    3.46 &    0.18 &  --1.49 &    0.44 &    0.03 && Ti \textsc{II} &  32 &    3.70 &    0.17 &  --1.25 &    0.40 &    0.03 \\
   V \textsc{I} &   6 &    1.88 &    0.08 &  --2.05 &  --0.12 &    0.03 &&   V \textsc{I} &   6 &    2.24 &    0.10 &  --1.69 &  --0.05 &    0.04 \\
  Cr \textsc{I} &  18 &    3.52 &    0.13 &  --2.12 &  --0.20 &    0.03 &&  Cr \textsc{I} &  15 &    3.78 &    0.07 &  --1.86 &  --0.22 &    0.02 \\
 Cr \textsc{II} &   2 &    3.83 &    0.03 &  --1.81 &    0.11 &    0.02 && Cr \textsc{II} &   1 &    4.14 & \nodata &  --1.50 &    0.14 & \nodata \\
  Mn \textsc{I} &   9 &    3.14 &    0.14 &  --2.29 &  --0.36 &    0.05 &&  Mn \textsc{I} &   7 &    3.44 &    0.15 &  --1.99 &  --0.35 &    0.06 \\
  Fe \textsc{I} & 100 &    5.57 &    0.12 &  --1.93 &    0.00 &    0.01 &&  Fe \textsc{I} &  83 &    5.85 &    0.12 &  --1.65 &    0.00 &    0.01 \\
 Fe \textsc{II} &  16 &    5.57 &    0.08 &  --1.93 &    0.00 &    0.02 && Fe \textsc{II} &  17 &    5.85 &    0.08 &  --1.65 &    0.00 &    0.02 \\
  Co \textsc{I} &   4 &    3.08 &    0.12 &  --1.91 &    0.01 &    0.06 &&  Co \textsc{I} &   6 &    3.33 &    0.13 &  --1.66 &  --0.01 &    0.05 \\
  Ni \textsc{I} &  24 &    4.24 &    0.12 &  --1.98 &  --0.05 &    0.03 &&  Ni \textsc{I} &  25 &    4.50 &    0.15 &  --1.72 &  --0.07 &    0.03 \\
  Cu \textsc{I} &   1 &    1.72 & \nodata &  --2.47 &  --0.54 & \nodata &&  Cu \textsc{I} &   1 &    2.15 & \nodata &  --2.04 &  --0.40 & \nodata \\
  Zn \textsc{I} &   2 &    2.65 &    0.10 &  --1.91 &    0.02 &    0.07 &&  Zn \textsc{I} &   2 &    2.79 &    0.04 &  --1.77 &  --0.13 &    0.03 \\
 Sr \textsc{II} &   1 &    0.90 & \nodata &  --1.97 &  --0.04 & \nodata && Sr \textsc{II} &   1 &    1.20 & \nodata &  --1.67 &  --0.03 & \nodata \\
  Y \textsc{II} &   3 &    0.12 &    0.10 &  --2.09 &  --0.16 &    0.06 &&  Y \textsc{II} &   3 &    0.35 &    0.10 &  --1.86 &  --0.21 &    0.06 \\
 Ba \textsc{II} &   2 &    0.19 &    0.08 &  --1.99 &  --0.06 &    0.06 && Ba \textsc{II} &   2 &    0.47 &    0.05 &  --1.71 &  --0.07 &    0.04 \\
 La \textsc{II} &   2 &  --0.73 &    0.03 &  --1.83 &    0.09 &    0.02 && La \textsc{II} &   2 &  --0.54 &    0.07 &  --1.65 &    0.00 &    0.05 \\
 Nd \textsc{II} &   8 &  --0.40 &    0.04 &  --1.82 &    0.10 &    0.02 && Nd \textsc{II} &   7 &  --0.18 &    0.07 &  --1.60 &    0.04 &    0.03 \\
 Eu \textsc{II} &   1 &  --1.12 & \nodata &  --1.64 &    0.29 & \nodata && Eu \textsc{II} &   1 &  --0.93 & \nodata &  --1.45 &    0.19 & \nodata \\

\cline{1-7} \cline{9-15} \\ \\
\multicolumn{7}{c}{\textbf{HD 142948}} && \multicolumn{7}{c}{\textbf{OSS 3}} \\
\cline{1-7} \cline{9-15}


   C (CH)       &   2 &    7.75 &    0.15 &  --0.68 &    0.03 &    0.15 &&   C (CH)       &   2 &    7.54 &    0.15 &  --0.89 &    0.00 &    0.15 \\
   O \textsc{I} &   2 &    8.57 &    0.02 &  --0.12 &    0.52 &    0.01 &&   O \textsc{I} &   1 & $<$8.41 & \nodata &$<$--0.27& $<$0.52 & \nodata \\
  Na \textsc{I} &   2 &    5.69 &    0.02 &  --0.55 &    0.09 &    0.01 &&  Na \textsc{I} &   2 &    5.22 &    0.01 &  --1.02 &  --0.16 &    0.01 \\
  Mg \textsc{I} &   6 &    7.34 &    0.10 &  --0.26 &    0.38 &    0.04 &&  Mg \textsc{I} &   6 &    6.70 &    0.14 &  --0.90 &  --0.04 &    0.06 \\
  Al \textsc{I} &   4 &    6.03 &    0.09 &  --0.42 &    0.22 &    0.04 &&  Al \textsc{I} &   2 &    5.27 &    0.11 &  --1.18 &  --0.32 &    0.08 \\
  Si \textsc{I} &   8 &    7.10 &    0.15 &  --0.41 &    0.23 &    0.05 &&  Si \textsc{I} &   8 &    6.72 &    0.16 &  --0.79 &    0.07 &    0.06 \\
   K \textsc{I} &   1 &    5.17 &    0.00 &    0.14 &    0.78 &    0.00 &&   K \textsc{I} &   1 &    4.85 & \nodata &  --0.18 &    0.68 & \nodata \\
  Ca \textsc{I} &  18 &    5.93 &    0.11 &  --0.41 &    0.23 &    0.03 &&  Ca \textsc{I} &  21 &    5.71 &    0.13 &  --0.63 &    0.23 &    0.03 \\
 Sc \textsc{II} &  13 &    2.81 &    0.09 &  --0.34 &    0.30 &    0.03 && Sc \textsc{II} &  11 &    2.24 &    0.09 &  --0.91 &  --0.05 &    0.03 \\
  Ti \textsc{I} &  22 &    4.44 &    0.10 &  --0.51 &    0.14 &    0.02 &&  Ti \textsc{I} &  19 &    4.17 &    0.15 &  --0.78 &    0.08 &    0.03 \\
 Ti \textsc{II} &  31 &    4.77 &    0.20 &  --0.18 &    0.47 &    0.04 && Ti \textsc{II} &  37 &    4.42 &    0.24 &  --0.53 &    0.33 &    0.04 \\
   V \textsc{I} &   4 &    3.44 &    0.09 &  --0.49 &    0.15 &    0.05 &&   V \textsc{I} &   2 &    2.97 &    0.05 &  --0.88 &  --0.02 &    0.04 \\
  Cr \textsc{I} &  11 &    4.82 &    0.05 &  --0.82 &  --0.17 &    0.01 &&  Cr \textsc{I} &  17 &    4.73 &    0.20 &  --0.91 &  --0.05 &    0.05 \\
 Cr \textsc{II} &   2 &    5.11 &    0.01 &  --0.53 &    0.11 &    0.00 && Cr \textsc{II} &   3 &    4.89 &    0.26 &  --0.75 &    0.11 &    0.15 \\
  Mn \textsc{I} &   8 &    4.74 &    0.18 &  --0.69 &  --0.05 &    0.06 &&  Mn \textsc{I} &   3 &    4.05 &    0.03 &  --1.38 &  --0.42 &    0.02 \\
  Fe \textsc{I} &  68 &    6.86 &    0.13 &  --0.64 &    0.00 &    0.02 &&  Fe \textsc{I} &  97 &    6.64 &    0.20 &  --0.86 &    0.00 &    0.02 \\
 Fe \textsc{II} &  16 &    6.87 &    0.14 &  --0.63 &    0.01 &    0.03 && Fe \textsc{II} &  27 &    6.64 &    0.24 &  --0.86 &    0.00 &    0.05 \\
  Co \textsc{I} &   5 &    4.49 &    0.08 &  --0.50 &    0.14 &    0.04 &&  Co \textsc{I} &   4 &    4.27 &    0.26 &  --0.72 &    0.14 &    0.13 \\
  Ni \textsc{I} &  21 &    5.62 &    0.16 &  --0.60 &    0.04 &    0.03 &&  Ni \textsc{I} &  25 &    5.33 &    0.24 &  --0.89 &  --0.03 &    0.05 \\
  Cu \textsc{I} &   1 &    3.90 & \nodata &  --0.29 &    0.35 & \nodata &&  Cu \textsc{I} &   1 &    2.53 & \nodata &  --1.66 &  --0.80 & \nodata \\
  Zn \textsc{I} &   2 &    4.13 &    0.00 &  --0.42 &    0.22 &    0.00 &&  Zn \textsc{I} &   2 &    3.67 &    0.15 &  --0.89 &  --0.03 &    0.10 \\
 Sr \textsc{II} &   1 &    2.40 & \nodata &  --0.47 &    0.17 & \nodata && Sr \textsc{II} &   1 &    1.95 & \nodata &  --0.92 &  --0.06 & \nodata \\
  Y \textsc{II} &   3 &    1.74 &    0.24 &  --0.47 &    0.17 &    0.14 &&  Y \textsc{II} &   2 &    1.28 &    0.35 &  --0.93 &  --0.07 &    0.25 \\
 Ba \textsc{II} &   2 &    1.44 &    0.02 &  --0.74 &  --0.09 &    0.01 && Ba \textsc{II} &   2 &    1.52 &    0.13 &  --0.66 &    0.20 &    0.09 \\
 La \textsc{II} &   2 &    0.61 &    0.10 &  --0.49 &    0.15 &    0.07 && La \textsc{II} &   1 & $<$0.57 & \nodata &$<$--1.53& $<$0.36 & \nodata \\
 Nd \textsc{II} &   4 &    1.02 &    0.07 &  --0.40 &    0.24 &    0.04 && Nd \textsc{II} &   1 &    1.46 & \nodata &    0.04 &    0.90 & \nodata \\
 Eu \textsc{II} &   1 &    0.25 & \nodata &  --0.27 &    0.37 & \nodata && Eu \textsc{II} &   1 &$<$--0.09& \nodata &$<$--0.61& $<$0.28 & \nodata \\

\cline{1-7} \cline{9-15} \\ \\
\multicolumn{7}{c}{\textbf{OSS 6}} && \multicolumn{7}{c}{\textbf{OSS 8}} \\
\cline{1-7} \cline{9-15}


   C (CH)       &   2 &    6.39 &    0.20 &  --2.04 &  --0.37 &    0.15 &&   C (CH)       &   2 &    6.85 &    0.20 &  --1.58 &    0.00 &    0.15 \\
   O \textsc{I} &   1 & $<$7.08 & \nodata &$<$--1.61& $<$0.06 & \nodata &&   O \textsc{I} &   1 & $<$7.46 & \nodata &$<$--1.23& $<$0.35 & \nodata \\
  Na \textsc{I} &   1 &    4.26 & \nodata &  --1.98 &  --0.23 & \nodata &&  Na \textsc{I} &   1 &    4.36 & \nodata &  --1.88 &  --0.26 & \nodata \\
  Mg \textsc{I} &   6 &    5.94 &    0.21 &  --1.66 &    0.09 &    0.09 &&  Mg \textsc{I} &   6 &    6.10 &    0.15 &  --1.49 &    0.13 &    0.06 \\
  Al \textsc{I} &   1 &    4.84 & \nodata &  --1.61 &    0.14 & \nodata &&  Al \textsc{I} &   1 & $<$5.21 & \nodata &$<$--1.24& $<$0.34 & \nodata \\
  Si \textsc{I} &   3 &    5.84 &    0.07 &  --1.67 &    0.09 &    0.04 &&  Si \textsc{I} &   2 &    6.01 &    0.13 &  --1.50 &    0.12 &    0.09 \\
   K \textsc{I} &   1 &    3.86 & \nodata &  --1.17 &    0.58 & \nodata &&   K \textsc{I} &   1 &    4.24 & \nodata &  --0.79 &    0.83 & \nodata \\
  Ca \textsc{I} &  21 &    4.78 &    0.15 &  --1.56 &    0.19 &    0.03 &&  Ca \textsc{I} &  22 &    5.03 &    0.14 &  --1.31 &    0.31 &    0.03 \\
 Sc \textsc{II} &   5 &    1.10 &    0.10 &  --2.05 &  --0.31 &    0.05 && Sc \textsc{II} &  16 &    1.47 &    0.15 &  --1.68 &  --0.06 &    0.04 \\
  Ti \textsc{I} &  23 &    3.09 &    0.18 &  --1.86 &  --0.11 &    0.04 &&  Ti \textsc{I} &  20 &    3.31 &    0.11 &  --1.64 &  --0.02 &    0.02 \\
 Ti \textsc{II} &  35 &    3.33 &    0.19 &  --1.62 &    0.13 &    0.03 && Ti \textsc{II} &  37 &    3.59 &    0.21 &  --1.36 &    0.26 &    0.03 \\
   V \textsc{I} &   2 &    1.74 &    0.12 &  --2.19 &  --0.44 &    0.08 &&   V \textsc{I} &   2 &    2.29 &    0.18 &  --1.64 &  --0.02 &    0.13 \\
  Cr \textsc{I} &  20 &    3.71 &    0.13 &  --1.93 &  --0.18 &    0.03 &&  Cr \textsc{I} &  17 &    3.82 &    0.12 &  --1.82 &  --0.19 &    0.03 \\
 Cr \textsc{II} &   2 &    4.10 &    0.06 &  --1.54 &    0.21 &    0.04 && Cr \textsc{II} &   3 &    4.23 &    0.04 &  --1.41 &    0.21 &    0.02 \\
  Mn \textsc{I} &   2 &    3.23 &    0.01 &  --2.20 &  --0.45 &    0.01 &&  Mn \textsc{I} &   5 &    3.53 &    0.18 &  --1.90 &  --0.28 &    0.08 \\
  Fe \textsc{I} &  66 &    5.75 &    0.11 &  --1.75 &    0.00 &    0.01 &&  Fe \textsc{I} &  60 &    5.88 &    0.15 &  --1.62 &    0.00 &    0.02 \\
 Fe \textsc{II} &  15 &    5.74 &    0.13 &  --1.76 &    0.00 &    0.03 && Fe \textsc{II} &  12 &    5.88 &    0.16 &  --1.62 &    0.00 &    0.05 \\
  Co \textsc{I} &   3 &    3.00 &    0.36 &  --1.99 &  --0.23 &    0.21 &&  Co \textsc{I} &   4 &    3.35 &    0.14 &  --1.64 &  --0.02 &    0.07 \\
  Ni \textsc{I} &  22 &    4.39 &    0.13 &  --1.83 &  --0.08 &    0.03 &&  Ni \textsc{I} &  20 &    4.52 &    0.15 &  --1.70 &  --0.08 &    0.03 \\
  Cu \textsc{I} &   1 &    1.02 & \nodata &  --3.17 &  --1.42 & \nodata &&  Cu \textsc{I} &   1 &    1.68 & \nodata &  --2.51 &  --0.89 & \nodata \\
  Zn \textsc{I} &   2 &    2.46 &    0.14 &  --2.09 &  --0.34 &    0.10 &&  Zn \textsc{I} &   1 &    2.89 & \nodata &  --1.67 &  --0.05 & \nodata \\
 Sr \textsc{II} &   1 &    0.73 & \nodata &  --2.14 &  --0.39 & \nodata && Sr \textsc{II} &   1 &    0.79 & \nodata &  --2.08 &  --0.46 & \nodata \\
  Y \textsc{II} &   1 &$<$--0.59& \nodata &$<$--2.77&$<$--1.10& \nodata &&  Y \textsc{II} &   1 &$<$--0.41& \nodata &$<$--2.60&$<$--1.00& \nodata \\
 Ba \textsc{II} &   1 &  --0.08 & \nodata &  --2.26 &  --0.50 & \nodata && Ba \textsc{II} &   2 &    0.04 &    0.20 &  --2.14 &  --0.51 &    0.14 \\
 La \textsc{II} &   1 &$<$--1.00& \nodata &$<$--1.82&$<$--0.35& \nodata && La \textsc{II} &   1 &  --0.12 & \nodata &  --1.22 &    0.40 & \nodata \\
 Nd \textsc{II} &   2 &  --0.71 &    0.08 &  --2.13 &  --0.38 &    0.06 && Nd \textsc{II} &   1 &    0.08 & \nodata &  --1.34 &    0.28 & \nodata \\
 Eu \textsc{II} &   1 &$<$--1.45& \nodata &$<$--1.97&$<$--0.30& \nodata && Eu \textsc{II} &   1 &$<$--0.62& \nodata &$<$--1.14& $<$0.44 & \nodata \\

\cline{1-7} \cline{9-15} \\ \\
\multicolumn{7}{c}{\textbf{OSS 14}} && \multicolumn{7}{c}{\textbf{OSS 18}} \\
\cline{1-7} \cline{9-15}

   C (CH)       &   2 &    5.85 &    0.20 &  --2.58 &    0.08 &    0.15 &&   C (CH)       &   1 &    7.57 &    0.20 &  --0.86 &  --0.10 &    0.15 \\
   O \textsc{I} &   1 & $<$6.40 & \nodata &$<$--2.09& $<$0.57 & \nodata &&   O \textsc{I} &   0 & \nodata & \nodata & \nodata & \nodata & \nodata \\
  Na \textsc{I} &   1 & $<$3.87 & \nodata &$<$--2.37& $<$0.29 & \nodata &&  Na \textsc{I} &   1 &    5.37 & \nodata &  --0.87 &  --0.25 & \nodata \\
  Mg \textsc{I} &   8 &    5.19 &    0.21 &  --2.41 &    0.25 &    0.07 &&  Mg \textsc{I} &   5 &    7.01 &    0.25 &  --0.59 &    0.03 &    0.11 \\
  Al \textsc{I} &   0 & \nodata & \nodata & \nodata & \nodata & \nodata &&  Al \textsc{I} &   1 & $<$5.66 & \nodata &$<$--0.79&$<$--0.03& \nodata \\
  Si \textsc{I} &   2 &    5.50 &    0.04 &  --2.00 &    0.66 &    0.02 &&  Si \textsc{I} &   5 &    6.98 &    0.28 &  --0.53 &    0.09 &    0.13 \\
   K \textsc{I} &   1 &    2.66 & \nodata &  --2.37 &    0.30 & \nodata &&   K \textsc{I} &   1 &    4.91 & \nodata &  --0.12 &    0.50 & \nodata \\
  Ca \textsc{I} &  16 &    3.89 &    0.15 &  --2.45 &    0.21 &    0.04 &&  Ca \textsc{I} &  19 &    5.82 &    0.14 &  --0.52 &    0.10 &    0.03 \\
 Sc \textsc{II} &   9 &    0.13 &    0.13 &  --3.02 &  --0.35 &    0.04 && Sc \textsc{II} &  11 &    2.64 &    0.18 &  --0.51 &    0.11 &    0.05 \\
  Ti \textsc{I} &  11 &    2.24 &    0.12 &  --2.71 &  --0.05 &    0.04 &&  Ti \textsc{I} &  17 &    4.27 &    0.24 &  --0.68 &  --0.06 &    0.06 \\
 Ti \textsc{II} &  32 &    2.37 &    0.18 &  --2.58 &    0.09 &    0.03 && Ti \textsc{II} &  25 &    4.44 &    0.33 &  --0.51 &    0.11 &    0.07 \\
   V \textsc{I} &   1 & $<$1.61 & \nodata &$<$--2.32& $<$0.34 & \nodata &&   V \textsc{I} &   4 &    3.26 &    0.01 &  --0.67 &  --0.05 &    0.00 \\
  Cr \textsc{I} &  10 &    2.54 &    0.23 &  --3.10 &  --0.43 &    0.07 &&  Cr \textsc{I} &  18 &    4.83 &    0.28 &  --0.81 &  --0.19 &    0.06 \\
 Cr \textsc{II} &   1 &    3.06 & \nodata &  --2.58 &    0.09 & \nodata && Cr \textsc{II} &   1 &    5.29 & \nodata &  --0.35 &    0.27 & \nodata \\
  Mn \textsc{I} &   4 &    2.20 &    0.20 &  --3.23 &  --0.56 &    0.10 &&  Mn \textsc{I} &   3 &    4.30 &    0.09 &  --1.13 &  --0.51 &    0.05 \\
  Fe \textsc{I} & 133 &    4.83 &    0.20 &  --2.67 &    0.00 &    0.02 &&  Fe \textsc{I} &  48 &    6.88 &    0.21 &  --0.62 &    0.00 &    0.03 \\
 Fe \textsc{II} &  16 &    4.83 &    0.16 &  --2.67 &  --0.01 &    0.04 && Fe \textsc{II} &  12 &    6.88 &    0.25 &  --0.62 &    0.00 &    0.07 \\
  Co \textsc{I} &   3 &    2.32 &    0.48 &  --2.67 &  --0.01 &    0.28 &&  Co \textsc{I} &   3 &    4.65 &    0.41 &  --0.34 &    0.28 &    0.24 \\
  Ni \textsc{I} &   5 &    3.51 &    0.07 &  --2.71 &  --0.04 &    0.03 &&  Ni \textsc{I} &  17 &    5.43 &    0.24 &  --0.79 &  --0.17 &    0.06 \\
  Cu \textsc{I} &   1 & $<$0.77 & \nodata &$<$--3.42&$<$--0.75& \nodata &&  Cu \textsc{I} &   1 &    3.14 & \nodata &  --1.05 &  --0.43 & \nodata \\
  Zn \textsc{I} &   1 & $<$1.92 & \nodata &$<$--2.64& $<$0.02 & \nodata &&  Zn \textsc{I} &   1 &    3.87 & \nodata &  --0.69 &  --0.07 & \nodata \\
 Sr \textsc{II} &   1 &  --0.22 & \nodata &  --3.09 &  --0.42 & \nodata && Sr \textsc{II} &   1 &    2.31 & \nodata &  --0.56 &    0.06 & \nodata \\
  Y \textsc{II} &   1 &$<$--0.86& \nodata &$<$--3.10&$<$--0.40& \nodata &&  Y \textsc{II} &   1 &    1.92 & \nodata &  --0.29 &    0.33 & \nodata \\
 Ba \textsc{II} &   2 &  --1.85 &    0.05 &  --4.04 &  --1.37 &    0.04 && Ba \textsc{II} &   2 &    1.76 &    0.05 &  --0.42 &    0.20 &    0.04 \\
 La \textsc{II} &   1 &$<$--0.91& \nodata &$<$--2.01& $<$0.65 & \nodata && La \textsc{II} &   1 &    1.29 & \nodata &    0.19 &    0.81 & \nodata \\
 Nd \textsc{II} &   1 &$<$--0.56& \nodata &$<$--1.98& $<$0.68 & \nodata && Nd \textsc{II} &   2 &    1.71 &    0.02 &    0.29 &    0.91 &    0.01 \\
 Eu \textsc{II} &   1 &$<$--1.14& \nodata &$<$--1.66& $<$1.00 & \nodata && Eu \textsc{II} &   1 &    0.72 & \nodata &    0.20 &    0.82 & \nodata \\

\end{longtable*}

\begin{deluxetable*}{lcccccc}
\tablecolumns{1}
\tabletypesize{\scriptsize}
\tablecaption{Abundance Uncertainties Due to Stellar Parameters\label{tab:chemical-abundance-uncertainties}}
\tablehead{
 & &  & & & \multicolumn{2}{c}{\textbf{Total Uncertainty}} \\
 \cline{6-7}
	\colhead{Species} &
	\colhead{$T_{\rm eff}+\sigma(T_{\rm eff})$} &
	\colhead{$\log{g}+\sigma(\log{g})$} &
	\colhead{${v_t}+\sigma(v_t)$} &
	\colhead{$Max(0.10, S.D.)/\sqrt(N)$} & 
	\colhead{[X/H]} &
	\colhead{[X/Fe]} \\
	 & $\Delta$abundance & $\Delta$abundance & $\Delta$abundance & (dex) & (dex) & (dex)
 }
\startdata
\\
\multicolumn{7}{c}{\textbf{OSS 3}} \\
\hline
 Na \textsc{I}   	&  +0.06	&  +0.00	& --0.01	& 0.07	& 0.09	& 0.08	\\
 Mg \textsc{I}   	&  +0.07	& --0.02	& --0.02	& 0.06	& 0.09	& 0.18	\\
 Al \textsc{I}   	&  +0.04	&  +0.00	&  +0.00	& 0.08	& 0.09	& 0.10	\\
 Si \textsc{I}   	&  +0.03	&  +0.01	& --0.01	& 0.06	& 0.07	& 0.08	\\
 K \textsc{I}   	&  +0.07	& --0.03	& --0.04	& 0.10	& 0.13	& 0.23	\\
 Ca \textsc{I}   	&  +0.07	& --0.02	& --0.03	& 0.03	& 0.08	& 0.17	\\
 Sc \textsc{II}   	&  +0.00	&  +0.03	& --0.03	& 0.03	& 0.05	& 0.19	\\
 Ti \textsc{I}   	&  +0.12	& --0.01	& --0.05	& 0.03	& 0.14	& 0.14	\\
 Ti \textsc{II}   	&  +0.01	&  +0.02	& --0.06	& 0.04	& 0.07	& 0.15	\\
 V \textsc{I}   	&  +0.12	&  +0.00	& --0.03	& 0.06	& 0.13	& 0.06	\\
 Cr \textsc{I}   	&  +0.11	& --0.01	& --0.06	& 0.05	& 0.13	& 0.13	\\
 Cr \textsc{II}   	& --0.02	&  +0.04	& --0.04	& 0.15	& 0.16	& 0.15	\\
 Mn \textsc{I}   	&  +0.10	& --0.02	& --0.03	& 0.11	& 0.16	& 0.21	\\
 Fe \textsc{I}   	&  +0.08	&  +0.00	& --0.04	& 0.02	& 0.09	&\nodata\\
 Fe \textsc{II}   	& --0.03	&  +0.04	& --0.03	& 0.05	& 0.07	&\nodata\\
 Co \textsc{I}   	&  +0.11	& --0.01	& --0.07	& 0.13	& 0.19	& 0.19	\\
 Ni \textsc{I}   	&  +0.07	&  +0.00	& --0.04	& 0.05	& 0.09	& 0.04	\\
 Cu \textsc{I}   	&  +0.10	&  +0.01	& --0.05	& 0.10	& 0.15	& 0.06	\\
 Zn \textsc{I}   	&  +0.00	&  +0.02	& --0.04	& 0.10	& 0.11	& 0.11	\\
 Sr \textsc{II}   	&  +0.03	&  +0.00	& --0.01	& 0.10	& 0.10	& 0.11	\\
 Y \textsc{II}   	&  +0.01	&  +0.03	& --0.02	& 0.25	& 0.25	& 0.16	\\
 Ba \textsc{II}   	&  +0.02	&  +0.01	& --0.06	& 0.09	& 0.11	& 0.05	\\
 Nd \textsc{II}   	&  +0.04	&  +0.04	& --0.04	& 0.10	& 0.12	& 0.21
\enddata
\tablenotetext{}{Table \ref{tab:chemical-abundance-uncertainties} is published for all program stars in the electronic edition. A portion is shown here for guidance regarding its form and content.}
\end{deluxetable*}

In order to investigate the abundance uncertainties due to stellar parameters, we have generated model atmospheres for a $\pm1\sigma$ offset in each stellar parameter (Section \ref{sec:stellar-parameter-uncertainties}; $T_{\rm eff}$, $\log{g}$, $\xi$) independently, and calculated the resultant mean abundance offset from our EWs. The resultant abundance changes were added in quadrature with $\sigma_{\bar{x}}$ to provide a total uncertainty in [X/H]. These total uncertainties are tabulated in Table \ref{tab:chemical-abundance-uncertainties} for all standard and program stars. While this yields an uncertainty for our abundance ratios in [X/H], we are often interested in how elemental abundances vary with respect to iron. We have calculated these uncertainties following \citet{johnson_2002}:
\begin{equation}
	\sigma^{2}\left({\rm A/B}\right) = \sigma^{2}\left({\rm A}\right) + \sigma^{2}\left({\rm B}\right) - 2\sigma_{{\rm A,B}}
\end{equation}

\noindent{}where the covariance between elements A and B ($\sigma_{{\rm A,B}}$) is given by
\small
\begin{multline}
	\sigma_{\rm A,B} = \left(\frac{\partial\log{\epsilon_{\rm A}}}{\partial T_{\rm eff}}\right)\left(\frac{\partial\log\epsilon_{\rm B}}{\partial T_{\rm eff}}\right)\sigma_{T_{\rm eff}}^2 \\
	+ \left(\frac{\partial\log{\epsilon_{\rm A}}}{\partial\log{g}}\right)\left(\frac{\partial\log\epsilon_{\rm B}}{\partial\log{g}}\right)\sigma_{\log{g}}^2	+ \left(\frac{\partial\log{\epsilon_{\rm A}}}{\partial\xi}\right)\left(\frac{\partial\log\epsilon_{\rm B}}{\partial\xi}\right)\sigma_{\xi}^2 \\
	+ \left[\left(\frac{\partial\log{\epsilon_{\rm A}}}{\partial T_{\rm eff}}\right)\left(\frac{\partial\log\epsilon_{\rm B}}{\partial\log{g}}\right) + \left(\frac{\partial\log{\epsilon_{\rm A}}}{\partial\log{g}}\right)\left(\frac{\partial\log\epsilon_{\rm B}}{\partial T_{\rm eff}}\right)\right]\sigma_{T_{\rm eff},\log{g}} \\
	+ \left[\left(\frac{\partial\log{\epsilon_{\rm A}}}{\partial\xi}\right)\left(\frac{\partial\log\epsilon_{\rm B}}{\partial\log{g}}\right) + \left(\frac{\partial\log{\epsilon_{\rm A}}}{\partial\log{g}}\right)\left(\frac{\partial\log\epsilon_{\rm B}}{\partial\xi}\right)\right]\sigma_{\xi,\log{g}}
\end{multline}
\normalsize
The covariance between effective temperature and surface gravity has been calculated by sampling about the effective temperature and performing an ionization balance for the adjusted temperature. The resultant covariance is given as:

\begin{equation}
	\sigma_{T_{\rm eff},\log{g}} = \frac{1}{N}\sum_{i=1}^{N}\left(T_{{\rm eff},i} - \overline{T_{\rm eff}}\right)\left(\log{g_i} - \overline{\log{g}}\right)
\end{equation}

\noindent{}Covariance between $\xi$ and $\log{g}$ is calculated in the same manner. The total abundance uncertainty with respect to H and Fe (e.g., $\sigma($[X/H]), $\sigma$([X/Fe])) for all species is given in Table \ref{tab:chemical-abundance-uncertainties}. These total uncertainties have been adopted in all figures throughout this text. Due to the cancellation of systematic effects, some abundance ratios with respect to iron have lower uncertainties than their quoted absolute abundance uncertainties.

\section{Discussion}
\label{sec:discussion}

\subsection{Stream Membership}
\label{sec:stream-membership}

Given the low surface brightness of the Orphan stream, separating true members from interlopers can be particularly challenging. Before inferring any properties of the undiscovered parent satellite from our sample, we must examine whether our targets are stars truly from the disruption of the Orphan stream progenitor.

When compared against the mean of our high priority targets, the velocities of the low and medium probability candidates are +6.2 and $-23.7$\,km s$^{-1}$ different, respectively. Given this region of the stream has a low intrinsic velocity dispersion \citep{casey;et-al_2013a}, the significantly lower velocity of the medium-probability target, OSS 18, is intriguing. Perhaps more concerning is that the low and medium probability candidates are markedly more metal-rich than the high-priority targets: [Fe/H] = $-0.86$ and $-0.62$\,dex for OSS 3 and OSS 18, a difference of $+0.45$ and $+$0.28\,dex from the low-resolution measurements, respectively. Spectroscopic studies suggest the stream is significantly more metal-poor than [Fe/H] $\sim -0.8$\,dex \citep{belokurov;et-al_2007,newberg;et-al_2010,casey;et-al_2013a,sesar;et-al_2013}, making the association between the Orphan stream and OSS 3 or OSS 18 tenuous. It is also worth noting that these targets are the farthest candidates ($|B_{\rm Orphan}| \sim 0.5$\,$^\circ$) from the best-fit Orphan stream orbital plane deduced by \citet{newberg;et-al_2010}. On the basis of the observables we deduce that the lower probability members, OSS 3 and OSS 18, are unlikely to be disrupted Orphan stream members.

The three high-probability targets (OSS 6, 8 and 14) have velocities within 2.4\,km s$^{-1}$ of each other, consistent with the Orphan stream velocity. These velocities from high-resolution spectra confirm the very low line-of-sight velocity dispersion in this part of the stream. The metallicities of these targets are also in reasonable agreement with those found from low-resolution spectroscopy, implying the candidates are $\sim$20\,kpc away -- at approximately the same distance as the Orphan stream. For the remainder of this discussion, we consider OSS 6, 8 and 14, to be true disrupted members of the Orphan stream. 

\subsection{Metallicity Distribution Function}
Although the sample size is small, the three Orphan stream members have a wide range in metallicity, ranging from [Fe/H] = $-1.58$ to as metal-poor [Fe/H] = $-2.66$. From three members not much can be said about the metallicity distribution, other than to say it is wide, and inconsistent with a mono-metallic population (e.g., a globular cluster). \citet{newberg;et-al_2010} and \citet{sesar;et-al_2013} found the stream to have a mean metallicity of [Fe/H] $= -2.1$ from BHB and RR Lyrae stars. According to the metallicity gradient along the stream reported by \citet{sesar;et-al_2013}, we might expect a higher mean metallicity than [Fe/H] $> -2.1$ in our sample closer to the celestial equator. We note that our metallicity spread of [Fe/H] = $-1.58$ to $-2.66$ is consistent with that of \citet{sesar;et-al_2013}: [Fe/H] = $-1.5$ to $-2.7$, and the somewhat wider spread found by \citet{newberg;et-al_2010}: [Fe/H] = $-1.3$ to $-3$.

A total metallicity spread as wide as $\sim$1\,dex, or a standard deviation of $\sigma({\rm [Fe/H]}) = 0.56$ \citep{casey;et-al_2013a}, is consistent with the stochastic chemical enrichment of the present-day dSph galaxies \citep{mateo_1998,kirby;et-al_2011}. This is also compatible with the constraints from the dynamical constraints placed inferred by the arc and width of the stream, which suggest the progenitor is a dark matter-dominated system.

\subsubsection{[$\alpha$/Fe] abundance ratios}
The abundance trends of individual elements track the star formation history of a cluster or galaxy. In the Milky Way, this is most evident by the evolution of $\alpha$-element abundances with increasing metallicity. The $\alpha$-elements are produced in massive stars before being ejected to the interstellar medium by Type II supernovae. At later times, Type Ia supernovae begin to contribute to the Galaxy's chemical enrichment. Type Ia supernovae expel iron (like Type II), but do not produce significant amounts of $\alpha$-elements. As such, a net decrease in [$\alpha$/Fe] is observed upon the onset of Type Ia supernovae. This inflexion occurs in the Milky Way near [Fe/H] $\sim -0.7$, decreasing towards Solar-like values.

\citet{tolstoy;et-al_2003} first noted that stars in the present-day dSph galaxies were separated in [$\alpha$/Fe] from the majority of the Milky Way stars. In the dSph galaxies, a significant contribution of Type Ia supernovae is observable in [$\alpha$/Fe] abundance ratios near [Fe/H] $> -1.7$\,dex, with the exception of Draco, which shows low [$\alpha$/Fe] at all metallicities. Some dwarf spheroidal stars also have a large range in [$\alpha$/Fe] values, increasing up to values seen in field stars.

The Orphan stream stars have low [$\alpha$/Fe] abundance ratios ($\sim$0.22) with respect to the Milky Way sample ($\sim$0.40\,dex). The low- and medium-probability members -- which we deem to be field interlopers -- also have low [$\alpha$/Fe] abundance ratios, but given their overall metallicity these $\alpha$-element ratios are consistent with the thick disk enrichment of the Milky Way. Since our Orphan stream members (i.e., the high-probability members OSS 6, 8 and 14) are metal-poor, we can be confident we are not confusing their low-$\alpha$ signature for the standard chemical enrichment of the Milky Way. 

We stress that [$\alpha$/Fe], on its own, should never be used as a litmus test for accretion on to the Milky Way. One must carefully select candidates that appear to be stream members based on all the observables, and then examine their detailed abundances in order to infer the chemical evolution of the disrupted host. We are already quite confident these stars are truly disrupted Orphan stream members, and as such, we can say that the low [$\alpha$/Fe] abundance ratios observed in the Orphan stream are consistent with the low $\alpha$-element enhancement of the present-day dSph galaxies.

\subsubsection{[Ba/Y]}
Differences in [Ba/Y] ratios between the present-day dSph stars and the Milky Way have been observed by a number of groups \citep{shetrone;et-al_2003,venn;et-al_2004}. A number of possible explanations have been proposed to explain this offset, including changes in SNe II yields -- or by changing the frequency of SNe II explosions via adjustments to the initial mass function -- or altering the influence of $\alpha$-rich freeze-out \citep[e.g., see][for a discussion]{venn;et-al_2004}.

\begin{figure}[h]
	\includegraphics[width=\columnwidth]{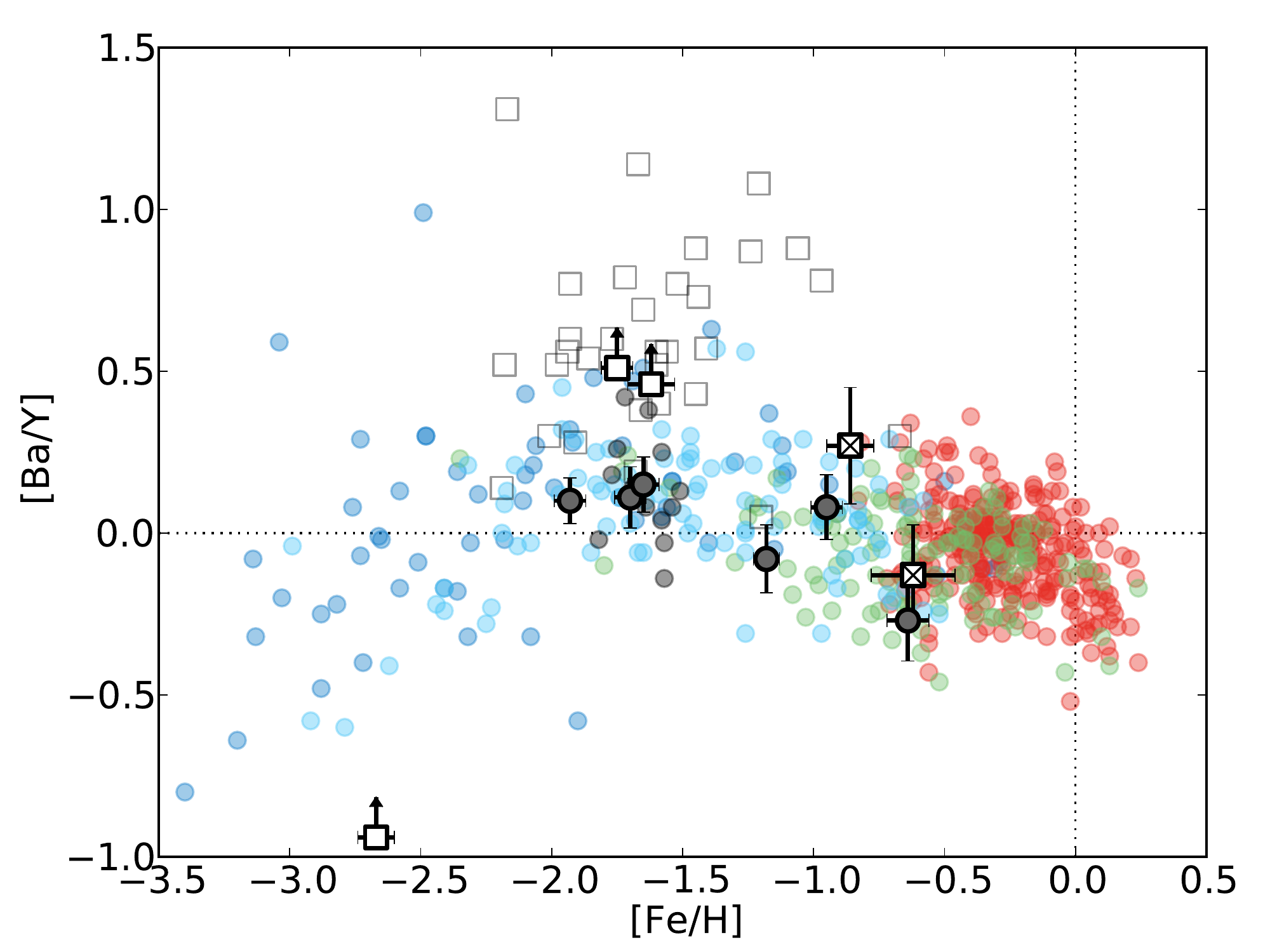}
	\caption{[Ba/Y] abundance ratios for the Orphan stream stars and field standards. Milky Way and dSph data compiled by \citet{venn;et-al_2004} is also shown. Markers and colours are the same described in Figure \ref{fig:alpha-fe}.}
	\label{fig:ba-y}
\end{figure}

Arbitrarily adjusting SNe II yields between two different nucleosynthesis sites appears a somewhat unlikely scenario. By employing SN frequency corrections in addition to using adjusted yields, \citet{qian;wasserburg} can reproduce the low [Ba/Y] trends in the Galaxy, but not the high [Ba/Y] ratios observed in the dSph galaxies. Leaving SNe II yields unchanged, these offsets could alternatively be reproduced by simply adjusting the frequency of SNe II events through the truncation of the upper initial mass function (IMF). In effect, massive stars would be less numerous, which are thought to be the primary production sites for the first $n$-capture peak elements (e.g., yttrium in this case). Because higher $\alpha$-element yields are also expected for massive stars, this may affect both the [Ba/Y] abundance and the overall [$\alpha$/Fe] abundance. However, whether the upper IMF differs between the dSph and the Milky Way remains an open question.


We have lower limits in [Ba/Y] for all Orphan stream stars. This is because Y was not detected in any of the stream stars. The standard stars and lower probability stream targets all have [Ba/Y] ratios that are consistent with the Milky Way for their given overall metallicity. This is shown in Figure \ref{fig:ba-y}. Although we have only lower limits for [Ba/Y], in OSS 6 and OSS 8 these are $\sim$+0.5\,dex offset from the main component of the Milky Way. This is a similar offset as that observed in the present-day dSphs, as compiled by \citet{venn;et-al_2004}. For the most metal-poor stream star, OSS 14, our limit on Y abundance yields a weak -- but consistent -- lower limit of [Ba/Y] $> -0.97$\,dex. We note that robust lower limits for [Ba/Y] in the most metal-poor dSph stars ([Fe/H]$ < -2$) are difficult to ascertain.

\subsection{Possible Parent Systems}
A number of associations have been proposed between the Orphan stream and known Milky Way satellites. However, most have been shown to be either unlikely or implausible. Here we discuss the possibility of association between the Orphan stream and NGC 2419, as well as Segue 1.

\subsubsection{NGC 2419}
The width and length of a tidal tail are a clear indication to the nature of the parent satellite. For the case of the Orphan stream, these characteristics favor a dark-matter dominated system (e.g., a dwarf galaxy). However, given the large apogalacticon of 90\,kpc for the Orphan stream, a relationship has been proposed between the Orphan stream and NGC 2419 \citep{bruns;kroupa_2011}. NGC 2419 is the most distant ($D \sim 85$\,kpc) -- and most luminous ($M_V \sim -9.6$) -- globular cluster in the Milky Way halo ($D > 20$\,kpc). Just like the Orphan stream, the system is also quite metal-poor: {[Fe/H] = --2.15}

Stars in globular clusters are known to exhibit peculiar chemical patterns, most notably in an anti-correlation between sodium and oxygen abundances \citep[e.g.,][]{carretta;et-al_2009a}. Following this, NGC 2419 is particularly special because it shows an unusual anti-correlation between magnesium and potassium, even stronger than the classical Na-O anti-correlation. This chemical signature is so far unique to NGC 2419, and illustrates the most extreme cases of Mg-depletion and K-enhancement seen anywhere in the Milky Way, or its satellite systems \citep{cohen;et-al_2011,mucciarelli;et-al_2012}.

\begin{figure}[h]
	\includegraphics[width=\columnwidth]{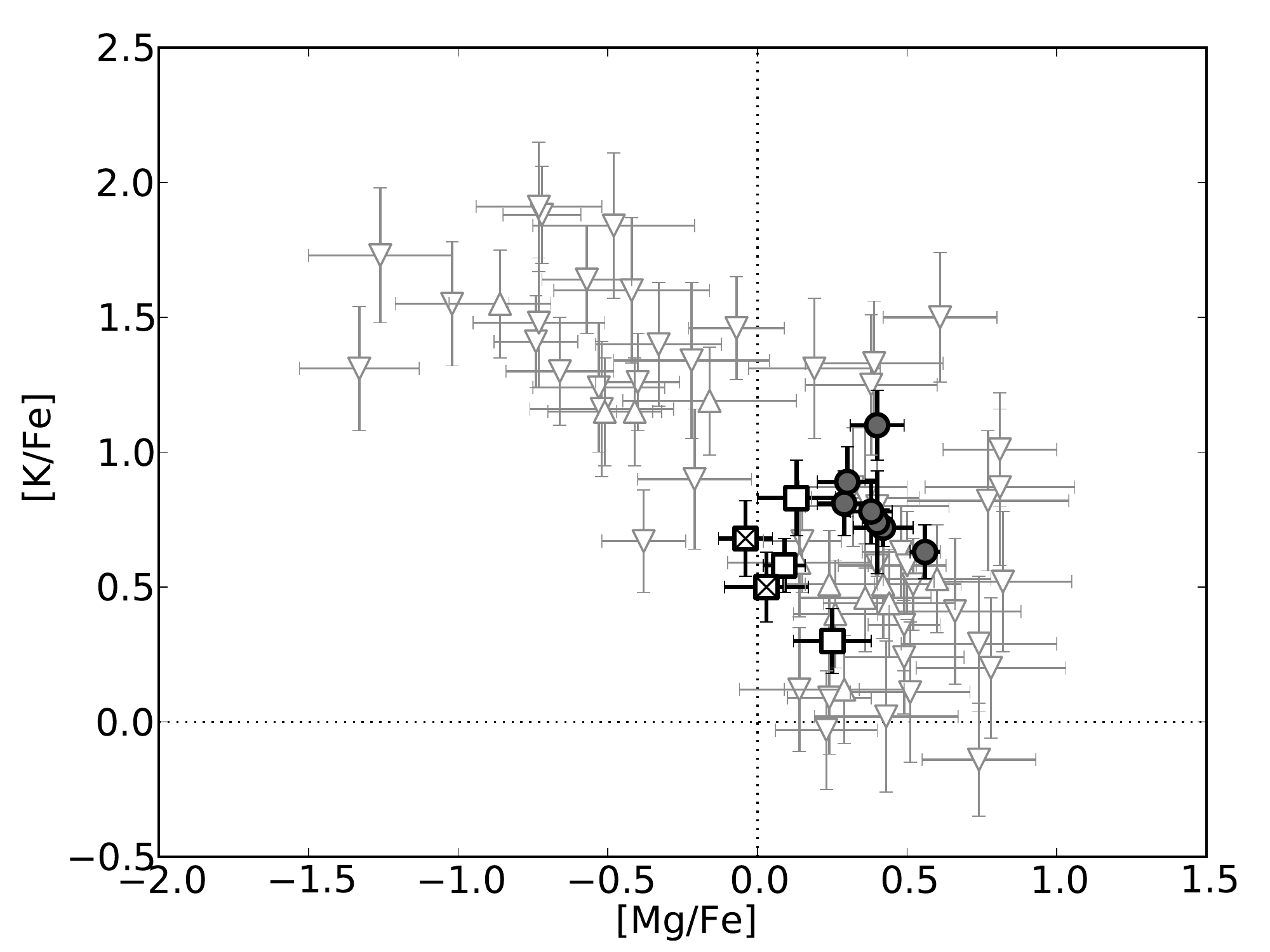}
	\caption{Magnesium and potassium abundance ratios for stars in the globular cluster NGC 2419 from \citet{cohen;et-al_2011} ($\vartriangle$) and \citet{mucciarelli;et-al_2012} ($\triangledown$). Standard stars and Orphan stream candidates are shown with the same markers used in Figure \ref{fig:alpha-fe}.}
	\label{fig:mg-k}
\end{figure}

We do not find any Orphan stream stars to be extremely depleted in [Mg/Fe] or enhanced in [K/Fe] (Figure \ref{fig:mg-k}), as found in NGC 2419 \citep{cohen;et-al_2010,cohen;et-al_2011,mucciarelli;et-al_2012}. If we consider two sub-populations in NGC 2419: a Mg-poor, K-rich and a Mg-normal, K-poor sample, then a two sample Kolmogorov-Smirnoff test demonstrates that we can be at least 99.7\% confident that our Orphan stream members are inconsistent with being drawn from the Mg-poor NGC 2419 population. We cannot make such statements about the Mg-normal population, since abundances for that sample are typical for the Milky Way. The Mg-poor and Mg-normal sub-populations each comprise about half of the observed NGC 2419 population \citep{mucciarelli;et-al_2012}. Given the uniqueness of the Mg-K anti-correlation, a single Orphan stream star with significant Mg-depletion or K-enhancement would have provided an exciting hint for the Orphan stream parent. However, given the stream's large metallicity spread \citep{casey;et-al_2013a,sesar;et-al_2013}, this indicates the stream is unlikely to be associated with NGC 2419 given the cluster's small dispersion in overall metallicity \citep[0.17\,dex;][]{mucciarelli;et-al_2012}. We therefore firmly rule out any association between NGC 2419 and the Orphan stream.



\subsubsection{Segue 1}


The only known satellite system with an ambiguous association with the Orphan stream is Segue 1. The system resides in a particularly crowded region of the sky: it lies extremely close to the perigalacton of the Orphan stream, and just south of the bifurcated Sagittarius stream. The original discovery of Segue 1 by \citet{belokurov;et-al_2007} suggested the system was an extended globular cluster, until recent studies showed the system was an ultrafaint dwarf galaxy \citep{geha;et-al_2009,norris;et-al_2010,simon;et-al_2011}. The similarities between Segue 1 and the Orphan stream are striking. In addition to being nearby on the sky, Segue 1's distance of 23\,kpc is consistent with the closest portion of the Orphan stream. The velocities between the two systems are also coincident: +114\,km s$^{-1}$ for Segue 1 and $\sim$+120\,km s$^{-1}$ for the nearest portion of the stream. Additionally, both systems exhibit extremely low velocity dispersions, on the order of 3-8\,km s$^{-1}$. However, stars linking the two systems have yet to be found. The association between the two is not obvious. As \citet{gilmore} summarises, `{\it are they just ships passing in the dark night?}'

The characteristics of the Orphan stream are clues to the nature of its progenitor: the arc length and intrinsic width suggests the parent system must be dark matter-dominated. Segue 1 is the most dark matter-dominated system known, such that it has become a prime focus for indirect dark matter detection experiments \citep{essig,baushev}. However, there are some doubts as to whether Segue 1 has experienced any tidal effects \citep{ostholt,simon;et-al_2011}. Segue 1 is well within the galactic tidal field, and initial indications of east-west tidal effects \citep{ostholt} seem to have been complicated by the crowded region bordering the system \citep{simon;et-al_2011}. Presently, no stars or tails clearly connecting the two systems have been found.

The chemistry of Segue 1 is particularly relevant here. Segue 1 hosts an extremely wide range in stellar metallicities: the mean is found to be [Fe/H] $= -2.5$ and the spread extends from [Fe/H] = $-3.4$ to $-1.6$\,dex \citep{simon;et-al_2011}. There is significant overlap between the metallicity distributions reported for Segue 1 and the Orphan stream. The high-resolution spectra in this study confirm the wide metallicity range of Orphan stream members found by \citet{casey;et-al_2013a}, and independently reported by \citet{sesar;et-al_2013} ([Fe/H] $= -1.5$ to $-2.7$). Thus, in addition to on-sky position, distances, velocity, and velocity dispersion, Segue 1 and the Orphan stream share a wide and consistent range in metallicities. In contrast to these observables, \citet{vargas;et-al_2013} and Frebel et al. (submitted) find Segue 1 members to be extremely $\alpha$-enhanced ([$\alpha$/Fe] $> 0.4$), unlike what we observe in disrupted Orphan stream members. 

We note that although the two systems appear to be related in some respect, if Segue 1 were the `parent' of the Orphan stream, the differing [$\alpha$/Fe] abundances and lack of tidal features surrounding Segue 1 -- after extensive examination -- is somewhat puzzling. Furthermore if Segue 1 is the disrupted host, and if the stream metallicity gradient presented by \citet{sesar;et-al_2013} is correct, that implies the stream becomes more metal-rich at greater distances from the parent system, contrary to what is typically observed. 

\section{Conclusions}
\label{sec:conclusions}

We present a chemical analysis of five Orphan stream candidates, three of which we confirm are true members from the disrupted Orphan stream parent satellite. The two non-members were both originally identified to have `low' or `medium' probability of stream membership from the low-resolution spectroscopy by \citet{casey;et-al_2013a}. We encourage high-resolution spectroscopic follow-up of the remaining high-probability Orphan stream members presented in \citet{casey;et-al_2013a}.

This work demonstrates the first detailed chemical study of the Orphan stream from high-resolution spectra. A large metallicity spread is present in the Orphan stream members, confirming the work by \citet{casey;et-al_2013a} and \citet{sesar;et-al_2013} that the Orphan stream is not mono-metallic. The spread in overall metallicity is consistent with the internal chemical evolution of the present-day dwarf galaxies. Detailed chemical abundances confirm this scenario. Low [$\alpha$/Fe] element ratios are observed in the stream stars, and lower limits of [Ba/Y] are ascertained, which sit well above the bulk component of the Milky Way for two of the three stream members. Thus, we present the first detailed chemical evidence that the parent of the Orphan stream is a dwarf galaxy.

On the basis of chemistry, we exclude the extended globular cluster NGC 2419 as a plausible parent to the Orphan stream. No firm link between Segue 1 and the Orphan stream has been identified. While the wide range in metallicities adds to the phase-space similarities between the two systems, the substantial difference in [$\alpha$/Fe] abundance ratios places doubt on any association. It appears the disrupting dwarf galaxy parent may reside in the southern sky, just waiting to be discovered.

\acknowledgements
A.R.C. is thankful to Martin Asplund, David Yong, Amanda Karakas, Ana Bonaca and Josh Simon for fruitful discussions and commentary on this work. Australian access to the Magellan Telescopes was supported through the National Collaborative Research Infrastructure Strategy of the Australian Federal Government. A.R.C. acknowledges the financial support through the Australian Research Council Laureate Fellowship LF0992131. S.K. and G.DaC. acknowledge the financial support from the Australian Research Council through Discovery Program DP120101237. A.F. acknowledges support from NSF grant AST-1255160.\\

\bibliographystyle{apj}

\end{document}